\documentclass[letterpaper, 10 pt, conference]{ieeeconf}

\IEEEoverridecommandlockouts                              % This command is only needed if 
\usepackage{setspace}
\usepackage{epsfig}  % for figures
\usepackage{epstopdf}
\usepackage{hyperref}

\usepackage{amssymb,amsmath,mathtools}
\usepackage{mathrsfs} 

\let\labelindent\relax
\usepackage{enumitem}
\usepackage{subfiles}

\usepackage{caption}
\usepackage{subcaption}
\makeatletter
\newcommand{\removelatexerror}{\let\@latex@error\@gobble}
\makeatother
\usepackage{xcolor}
\usepackage[ruled, vlined]{algorithm2e}

\SetCommentSty{mycommfont}

\SetAlFnt{\small}

\usepackage{lipsum}
\usepackage{cite}
\usepackage{thmtools,thm-restate}
\usepackage{multicol}
\usepackage{multirow}
\usepackage{bm}
\newtheorem{theorem}{Theorem}

\newtheorem{definition}{Definition}
\newtheorem{example}{Example}

\newcommand{\specificthanks}[1]{\@fnsymbol{#1}}% Inserts a specific \thanks symbol
\newcommand{\vertiii}[1]{{\left\vert\kern-0.25ex\left\vert\kern-0.25ex\left\vert #1 
    \right\vert\kern-0.25ex\right\vert\kern-0.25ex\right\vert}}

\newcommand{\HJ}[1]{{\color{black}{#1}}}
\begin{document}

\title{\LARGE \bf
	 Robust Environmental Mapping by Mobile Sensor Networks
}

\author{Hyongju~Park, Jinsun~Liu, Matthew~Johnson-Roberson and Ram~Vasudevan
	\thanks{*This work was supported by Ford Motor Company}
\thanks{Hyongju Park, Jinsun Liu, Matthew Johnson-Roberson, and Ram Vasudevan are at the University of Michigan, Ann Arbor, MI, 48109 USA {\texttt{\{hjcpark,jinsunl,mattjr,ramv\}@umich.edu}.}
}
}
\maketitle
\begin{abstract}
Constructing a spatial map of environmental parameters is a crucial step to preventing hazardous chemical leakages, forest fires, or while estimating a spatially distributed physical quantities such as terrain elevation. 
Although prior methods can do such mapping tasks efficiently via dispatching a group of autonomous agents, they are unable to ensure satisfactory convergence to the underlying ground truth distribution in a decentralized manner when any of the agents fail.
Since the types of agents utilized to perform such mapping are typically inexpensive and prone to failure, this results in poor overall mapping performance in real-world applications, which can in certain cases endanger human safety.
This paper presents a Bayesian approach for robust spatial mapping of environmental parameters by deploying a group of mobile robots capable of ad-hoc communication equipped with short-range sensors in the presence of hardware failures. 
Our approach first utilizes a variant of the Voronoi diagram to partition the region to be mapped into disjoint regions that are each associated with at least one robot.
These robots are then deployed in a decentralized manner to maximize the likelihood that at least one robot detects every target in their associated region despite a non-zero probability of failure.
A suite of simulation results is presented to demonstrate the effectiveness and robustness of the proposed method when compared to existing techniques.

\end{abstract}

%\begin{IEEEkeywords}
%multi-robot system, fault-tolerant algorithm, order$-k$ Voronoi diagram, coverage control, deployment.
%\end{IEEEkeywords}
%

% \fi
\IEEEpeerreviewmaketitle

\section{Introduction}
\label{sec:sec1}
%What is the problem?
%Why is it interesting and important?
%Why is it hard? (E.g., why do naive approaches fail?)
%Why hasn't it been solved before? (Or, what's wrong with previous proposed solutions? How does mine differ?)
%What are the key components of my approach and results? Also include any specific limitations.

%\subsection{Objective of this paper}

This paper studies environmental mapping via a team of mobile robots equipped with ad-hoc communication and sensing devices which we refer to as a \emph{Mobile Sensor Network} (MSN).
In particular, this paper focuses on the challenge of trying to estimate some unknown, spatially distributed target of interest given some \textit{a priori} measurements under the assumption that each robot in this network has limited sensing/processing capabilities.
MSNs have been an especially popular tool to perform environmental mapping due to their inexpensiveness which enables large-scale deployments \cite{connor2016airborne,schwager2017multi,cortez2011information,pahlajani2014networked,julian2012distributed,lynch2008decentralized}; however, this economical price-point betrays their susceptibility to hardware failures such as erroneous sensor readings.
This paper aims to develop a class of cooperative detection and deployment strategies that enable MSNs to autonomously and collectively obtain an accurate representation of an arbitrary environmental map efficiently while certifying robustness to a bounded number of sensor failures.

\begin{figure}
    \hspace*{-0.25cm}
    \includegraphics[width=3.4in,height=3.4in,keepaspectratio=false]{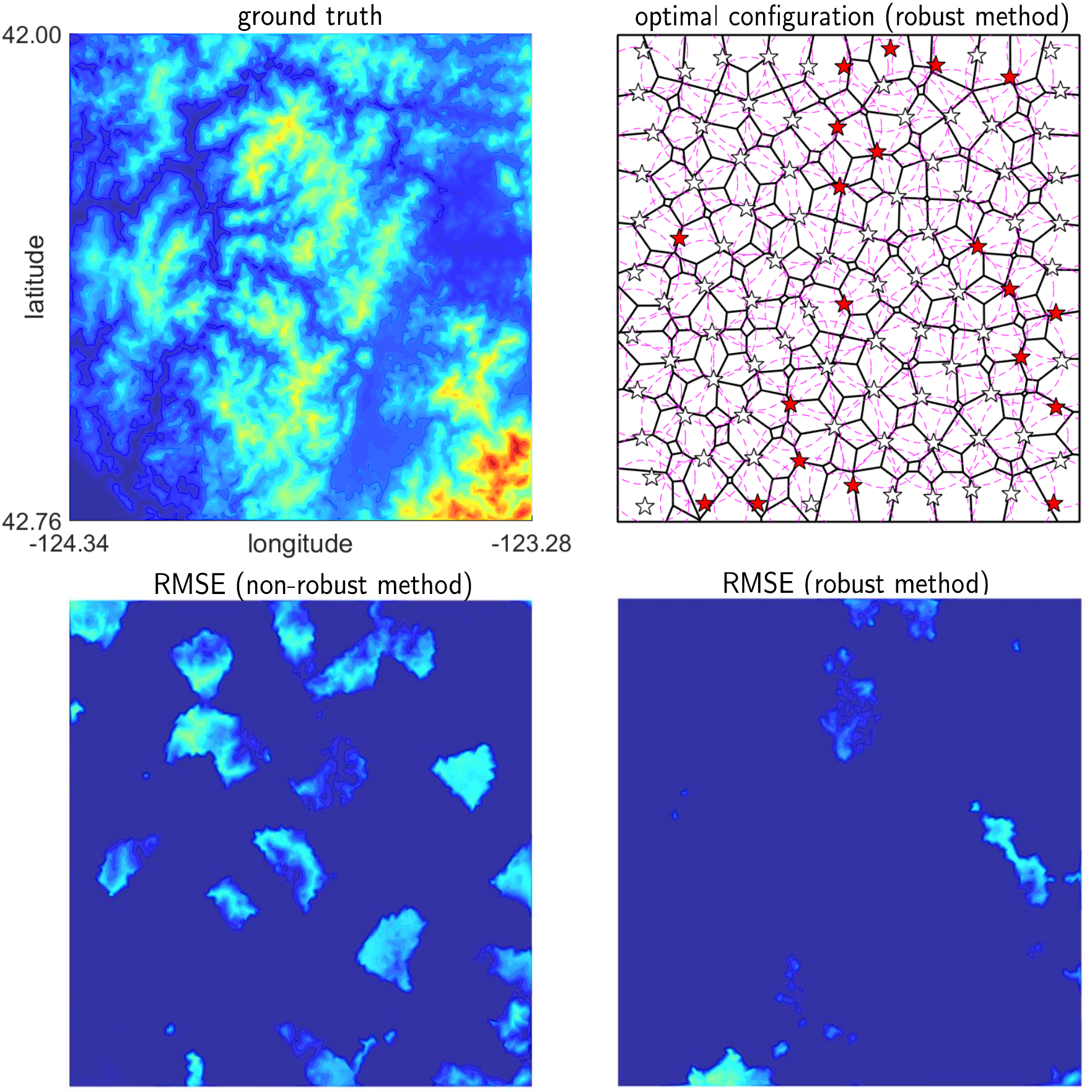}
    \caption{An illustration of the application of the robust, mobile sensor network deployment and reconstruction algorithm developed in this paper  (top right, robots are shown in stars, robots with failed sensors are shown in red, sensor footprints are drawn in dotted lines, and the partition is drawn as polygons) when applied to $100$ robots that are trying to build an elevation map of a mountainous region in Oregon USA (top left) when $20$ sensors have failed (red stars). 
    Despite the sensor failures, the root mean square error of the reconstructed map when compared to the ground truth using the method presented in this paper (bottom right) is significantly better than existing methods (bottom left).}
    \label{fig:fig9}
    \vspace*{-0.8cm}
\end{figure}

%\subsection{Related works}
%\subsubsection{Environmental mapping using mutual information gathering}
%[ISSUE \#1: COMPUTATIONAL COMPLEXITY, SCALABILITY TO MSNs ON ENVIRONMENTAL MAPPING TASKS]
%Others have employed a non-Bayesian method to perform target distribution mapping using a single mobile robot

Few methods have been proposed to accurately perform environmental mapping using a large number of mobile robots that can guarantee robustness to hardware failures while making realistic assumptions about a MSN.
For example, one of the most popular methods for addressing environmental mapping via MSNs has utilized the notion of mutual information to design controllers that follow an information gradient \cite{cortez2011information,pahlajani2014networked,schwager2017multi,julian2012distributed}. 
These approaches focus on linear dynamics and Gaussian noise models. 
Recently this technique was utilized to enable MSNs to estimate a map of finite events in the environments while avoiding probabilistic failures that arose due to nearby encounters with unknown hazards \cite{schwager2017multi}.
The computational complexity for computing this information gradient is exponential in the number of robots, sensor measurements, and environmental discretization cells \cite{schwager2017multi, julian2012distributed}.
More problematically, the computation of the gradient requires that every robot be omniscient, i.e., have current knowledge of every other robot’s position and sensor measurements.
For this reason, mutual information-based methods are generally restricted to small groups of robots with fully connected communication networks which has limited their potential real-world application. 

To overcome this computational complexity related problem, others have focused on devising relaxed techniques to perform information gathering.
For instance, some have proposed a fully decentralized strategy where the gradient of mutual information is used to drive a network of robots to perform environmental mapping \cite{julian2012distributed}.
To improve computational efficiency they relied on a sampling technique; however this restricted their ability to perform mapping of a general complex environment instead they focus on cell environments. 
%At least from the current stand, their approach has not been shown to be able to do mapping of a general complex environment, which is of our problem of interest. 
%Their experimental results presents the case of robot fleets are inferring finite number of cell environment. 
Others have tried to develop particle filter based techniques to enable the application of nonlinear and non-Gaussian target state and sensor models while approximating the mutual information \cite{hoffman2010mobile}. 
This method is shown to localize a target efficiently.
However this approach still assumes the existence of a centralized algorithm to fuse together the information from multiple sensors.

Rather than rely on the information gradient, others have employed algorithms that use information diffusion through communication network for environmental modeling \cite{lynch2008decentralized}.  
By utilizing the Average Consensus filter to share information among the robots in the network, this approach is scalable to large numbers of agents, is fully decentralized, and can even work under a switching network topology as long as the network is connected; however the approach is not spatially distributed and requires an additional connectivity maintenance algorithm \cite{yang2010decentralized} to ensure its convergence.

%
%however, 
%since the focus is on information diffusion rather than building a spatially distributed map, this approach may not be ideal for environmental mapping \Ram{also is this algorithm robust?}. 
%\Ram{I don't understand the relevance of these next to sentences...} If configured properly\footnote{One popular example of this is the work of \cite{cortes_coverage_2004} where centroidal Voronoi configuration of MSNs are proven to  achieves optimal coverage.}, 
%MSNs are expected to handle large region environmental mapping task efficiently while maintaining minimal overlapping sensing regions.

%\subsubsection{Approaches maximizing mutual information gain}

%The particle filtering was used in this paper not due to the complexity of our algorithm; but for propagating of belief over a continues target space.

%\textbf{ISSUE \#2: NO ROBUSTNESS GUARANTEE UNDER SENSOR FAILURE}

This paper presents a class of computationally efficient, scalable, decentralized deployment strategy that is robust to sensor failures. 
We employ classical higher order Voronoi tessellation  \cite{shamos1975closest} to achieve a spatially distributed allocation of MSNs for efficient environmental mapping.
In particular each region from the partition is assigned to multiple robots to provide robustness to sensor failures.
Although others have employed ordinary Voronoi tessellation for robot-target assignment towards efficient information gathering \cite{cortez2011information,bandyopadhyay2003energy,patten2013large,kemna2017multi}, these approaches are not guaranteed to converge to an underlying distribution in the case of even a single sensor failure \cite{hutchinson_robust_2012}.
To best of our knowledge, almost all studies about environmental mapping by MSNs have not take into account such adversarial scenarios, nor presented performance guarantees in terms of convergence to some ground truth value.
In addition, we consider a broad class of sensor failures which are not restricted to just failures associated with proximity to a hazard. Our cost function is the likelihood that the MSN will fail to make reliable measurement of the spatially distributed environmental parameters.
%formulated based on a particular spatial tessellation known as a higher order Voronoi partition \cite{shamos1975closest}.
We use gradient descent on the cost to design a decentralized deployment strategy for the MSN. 
By doing so, each robot can compute its gradient using merely local information without requiring communication with a central server.
In this paper, a central entity is only required to fuse and update the information gathered from MSNs, but not to generate control policies for robots as in typical mutual information gathering approaches \cite{schwager2017multi,julian2012distributed}. 
To generate an estimate for the underlying target distribution of the environment, this paper employs a particle filter with low discrepancy sampling. 

%Furthermore, our paper also employed particle filter along with low discrepancy sampling method for the purpose of approximating the belief of the complex, continuously distributed target rapidly.

In addition, this paper presents a novel combined sensor model that assigns different weights to robots by taking into account the spatial relationship between robots and a target state. 
This detection model is based on a classical binary model that depends on the configuration of robots \cite{viswanathan1997distributed,djuric2008target}. 
To connect the detection model to the measurement model we rely on a nonrestrictive assumption that if a robot fails to discern one target from another, it may not provide the correct sensor reading for the target. 
This assumption is similar to one used in a previous approach that also built a combined sensor model that was experimentally verified with the laser range finder and a panoramic camera measurement \cite{anguelov2004detecting}. 
This sensor model enables one to decouple the information state from the detection task, which can make computing the gradient computationally sound with a complexity that is linear with respect to the number of sensors. 

The main contributions of this paper are three-fold:
First, we adopt a higher order Voronoi tessellation for optimal robots-to-target assignment to provide robustness under a general class of sensor failures whose number is bounded.
Second, we present a novel sensor model, to remove the computational burden of maintaining mutual information in MSNs by decoupling information gathering and detection, while ensuring satisfactory mapping performance.
Finally, we propose a scalable, spatially distributed, computationally efficient, decentralized controller for MSNs which can perform environmental mapping task rapidly.
% All of which has been extensively validated via a suite of numerical simulations. 

\textit{Organization:}
The rest of the paper is organized as follows. 
Section \ref{sec:sec2} presents notation used in the remainder of the paper, formally defines the problem of interest.
Section \ref{sec:sec3} presents our combined probabilistic sensor model, and the deployment strategy is formally presented in Section \ref{sec:sec4}. 
Section \ref{sec:sec6} discusses an approximate belief update method via particle filters. 
The robustness of our deployment and effectiveness of the belief update approach is evaluated via numerical simulations in Sections \ref{sec:sec7}. Finally, Section \ref{sec:sec9} concludes the paper.
%
%In this paper, we propose a class of deployment strategies for MSNs aimed for robust environmental monitoring. 
%Instead of choosing control to be directly maximizing information gain (entropy), which has been known to be notoriously computationally demanding and has no known robustness properties to failures; we pose the a constrained optimization problem and use classical spatial tessellation to derive solution model. Due to complexity of the problem, we propose gradient descent algorithm and show that state propagated by the control will mitigate the effect of unreliable measurement to be equally used to propagate the belief, 
%Deployment through the control law is extremely useful to efficiently consider observations made at different locations using different robots.

%\section{Preliminaries}
\section{Problem Description}
\label{sec:sec2}

This section presents the notation used throughout the paper, an illustrative example, and the problem of interest.
%\subsection{Our system}
\subsection{Notations and Our System Definition}
\label{sec:sec21}
The italic bold font is used to describe random quantities, a subscript $t$ indicates that the value is measured at time step $t$, and $\mathbb{Z}_{\geq 0}$ denotes non-negative integers. %We may also suppress subscript $t$ if it is clear from the context. 
Given a continuous random variable $\bm{x}$, if it is distributed according to a Probability Density Function (PDF), we denote it by $f_{\bm{x}}$.
Given a discrete random variable $\bm{y}$, if it is distributed according to a Probability Mass Function (PMF), we denote it by $p_{\bm{y}}$.
Consider a group of $m$ mobile robots deployed in a workspace, i.e., ambient space, $\mathcal{Q} \subseteq \mathbb{R}^d$ where $d = 2,3$. 
This paper assumes $d=2$ though the presented framework generalizes to $d=3$. 
Let $\mathbb{S}^{d-1} = \lbrace s\in \mathbb{R}^d\mid \left\| s \right\|=1 \rbrace$ be a unit circle/sphere, then the state of $m$ robots is the set of locations and orientations at time $t$, and it is represented as an $m$-tuple $x_t = (x_t^1,\dots,x_t^m)$, where $x_t^i \in \mathcal{Q} \times \mathbb{S}^{d-1}$. The state of robots are assumed completely \emph{known}.
We denote by the set ${x}_{0:t}\coloneqq \lbrace {x}_0,\dots,{x}_t \rbrace$ the robot states up to time $t$. 

% The group of robots follow a way-point-based, deterministic motion model.
%
%
%Given a pair of states $(x_t,x_{t+1})$, robots follow a way-point-based, continuous-time, deterministic motion model with dynamic constraints:
% \begin{equation}
% x_{t+1} = x_t + u_t, \,\,\,\,\,\,\, t \in \mathbb{Z}_{\geq 0}
% \label{eq:constraint}
% \end{equation}
We assume that subsequent states satisfy some controlled dynamical system where $u_t \in \mathcal{U} \subseteq \mathbb{R}^d$ is the control that takes a system from $x_t$ to $x_{t+1}$.
%Let $u_t^{\star}$ be the \emph{optimal control policy}\footnote{In the current context, the optimal control policy is a sequence of control inputs (in discrete time domain) or control path (in continuous time domain) governed by dynamics of the vehicle. On example of such optimal control policy can be generated by Linear-Quadratic Regulator (LQR) if the dynamics were linear.} which drives robots' state from $x_t$ to $x_{t+1}$ in minimum time under the dynamic (or kinematic) constraints. 
% Let $u_{0:t}$ be the sequence of control policies up to time $t$.
We define a \emph{target} to be a physical object or some measurable quantity that is spatially distributed over a bounded domain.
Let $\bm{z}$ be the random vector representing \emph{target state} which consists of locations, $\bm{q} \in \mathcal{Q}$, and an information state (i.e., quantitative information about the target), $\bm{I} \in \mathcal{I} \subseteq \mathbb{R}$ where we let $\mathcal{I} =[I_{\min},\,I_{\max}]$.
We define $\mathcal{Z} = \mathcal{Q} \times \mathcal{I}$ as the \emph{target state space}.
Finally, we let $\mathcal{F} \subsetneq \lbrace 1,\dots,m \rbrace$ be the index set of robots whose sensors have failed.

\begin{example}[Airborne LIDARs for DEM generation]
Consider a group of autonomous aerial vehicles trying to acquire an accurate Digital Elevation Model (DEM)\footnote{A digital elevation model (DEM) is a digital 3D model of a terrain's surface created from terrain elevation data.} of some bounded region using airborne LIDAR measurements.
Suppose the state of the robot is some 2D location at some fixed height above the terrain. 
In this instance, the target state space would be made up of a longitude and latitude, $q \in \mathcal{Q}$ and an elevation at that point which belongs to the set $\mathcal{I} \subset \mathbb{R}$.
This paper explores how to determine the best way to deploy a finite number of agents to minimize the probability that they fail to detect a set of targets dispersed over a region.
Unfortunately, the LIDAR measurements from a part of the fleet may be corrupt or unreliable.
To ensure that we can guarantee the optimal target detection performance in adversarial scenarios, we develop a robust deployment strategy.
Subsequently, this paper explores how to efficiently reconstruct the terrain map (i.e. the distribution over the target state space) from a set of deployed robots.

% In this instance a detection, $\bm{y^i|z}_{t}$ means that robot $i$ at time $t$ received a LIDAR measurement from target state $\bm{z}$.
% Finally, 
% The first problem is: given a finite number of vehicles, how to deploy those vehicles such that the probability that robots will fail to targets dispersed over the region of interest stays minimum. 
% Taking such adversarial scenarios into account, it is crucial to design a deployment strategy along with an underlying sensor model, which can guarantee worst-case optimal target detection performance. 
% Finally, it is also important to process the measurement to update the belief on target state over time in efficient manner such that, within a reasonable time, the autonomous fleet can obtain an accurate DEM model.
\end{example}

%\subsection{An illustrative Example: Airborne LiDAR for DEM generation}
%\label{sec:sec22}

\subsection{Robust Deployment Strategy}
\label{sec:sec23}
%Let $\bm{y}_{D}$ a binary random tuple of size $m$ where each $\bm{y}_{D}^i \in \lbrace 0,\,1 \rbrace$ which returns the detection by $i$th robot.
Suppose we are given $z \in \mathcal{Z}$, then we let $\bm{y_{t}}|z=(\bm{y}_{t}^{1}|z,\dots,\bm{y}_{t}^m|z)$ be a binary random $m$-tuple which indicates whether an observation is made by $m$ robots at time step $t$ at a given target state $z$, where $\bm{y}_{t}^{i}|z \in \lbrace 0,\,1 \rbrace$ for each $i$.
% For the sake of convenience throughout the remainder of the document, we suppress this dependence on $z$ when we refer to $\bm{y}_{t}|z$, since it is usually clear in context.
Let the set $\bm{y}_{1:t}|z\coloneqq \lbrace \bm{y}_1|z,\dots,\bm{y}_t|z \rbrace$ denote the observations made by robots up to time $t$. 
For a given $q \in \mathcal{Q}$ and robot state $x_t \in \mathcal{Q}\times \mathbb{S}^1$, let $\bm{y}_{D,t}|x_t,q:=(\bm{y}_{D,t}^1|x_t,q,\dots,\bm{y}_{D,t}^m|x_t,q)$ be a binary random $m$-tuple which indicates whether $m$ robots with state $x_t$ are able to detect a target located at $q$.
We let $\bm{y}_{D,t}^i|x_t,q \in \lbrace 0,\,1 \rbrace$ for each $i$. 
We let $p(\bm{y}_{D,t}=\bm{0} \mid x_t, \bm{q}=q)$ be the probability of a joint event that a group of $m$ robots with location and orientation given by $x_t$ fail to detect a target at time $t$ if it is located at $q$, where $\bm{0}$ is an $m$-tuple of zeros. 
%WRITE EQUATIONS HERE....
For the case when each sensor belonging to the index set $\mathcal{F}$ has failed at time $t$, our aim is to find an optimal configuration that solves:
\begin{align}
& \min_{x_t \in \mathcal{Q}^m} 
p(\bm{y}_{D,t} = \bm{0}\mid x_t) \nonumber \\
& \,\,\,\,\,\,\textup{s.t. } \,\,\,p(\bm{y}_{D,t} = \bm{0} \mid x_t,\bm{q}=q,\mathcal{F}) <1, \,\,\,\,\,\,\, \forall q \in \mathcal{Q},
\label{mainprob}
\end{align}
%\begin{equation}
%\label{const}
%\end{equation}
where a given $\mathcal{F}$ and a target, there must be at least one robot that is able to make a reliable measurement. %, which is an essential constraint for the environmental mapping task.
%\footnote{This paper considers an average-case optimal solution with respect to sensor failures. For our purpose, the worst-case optimal (minimax) solution can, in fact, be too conservative.}.
%The target distribution of $\bm{q}$ has been marginalized out. 
Unfortunately, obtaining the global solution to this problem is proven to be NP-Hard by reduction from the simpler static ``locational'' optimization problem, the \emph{$m$-median problem}\footnote{The $m$-median problem is one of the popular locational optimization problems where the objective is to locate $m$ facilities to minimize the distance between demands and the facilities given some uniform prior. The problem is NP-Hard for a general graph (not necessarily a tree).}.
To overcome the computational complexity, we apply a gradient descent approach where the control policy at each time step minimizes the missed-detection probability of targets by robots at their future locations (one-step look-ahead). 
We utilize the higher order Voronoi tessellation for robot target assignment, which guarantees that the solution to \eqref{mainprob} is found under such an assignment \cite{shamos1975closest}.

%We will present a gradient decent method to derive a local solution to the problem given a constant probability of failure for all robots, i.e., $p(i \in \mathcal{F}) = p_{\mathcal{F}}^i$.
%Thus, given a target $q$ distributed over $\mathcal{Q}$, with prior  distribution $f_q$, 

\subsection{Combined Sensor Model}
\label{sec:sec24}
We assume that  a sensor can correctly measure a given target only if the sensor can detect the target \emph{a priori}.
We further assume that, if a sensor can detect a target, a measurement of the target by the sensor may be corrupted by noise.
% Whereas, missed-detections may lead to an unreliable measurement, which should not be considered as a correct one.  
These assumptions have been experimentally validated for example on a mobile robot that uses a laser range finder and a panoramic camera measurement \cite{anguelov2004detecting}.

% These assumptions have turned out to be non-restrictive according to the work of \cite{} in which the authors has experimentally verified a similar combined sensor model on a mobile robot using laser range finder and a panoramic camera measurement.

\subsection{Evaluation of Mapping Performance}
\label{sec:sec25}
We derive a particle filter to recursively update approximate beliefs on a particular unknown environment. 
Let $\hat{b}_t$ represent the approximate posterior probability distribution of the target state at time $t \in \mathbb{Z}_{\geq 0}$, the initial belief $\hat{b}_0$ is assumed to be a uniform density if no prior information on the target is available.
We let $\hat{b}^{\star}$ be the PMF estimate of the \emph{true posterior belief}\footnote{We shall assume, for the sake of discussion, that the true posterior target distribution can be obtained, e.g., via exhaustive search and measurements made by a MSN.}.
To this end, we quantify the difference between the true posterior belief, $b^{\star}$ and our method via the Kullback-Leibler (K-L) divergence.
%\footnote{K-L divergence is often used to measure the difference between two probability distributions; given $f_{\bm{x}},f_{\bm{y}}$,		$D_{\text{KL}} \left( \bm{x} \middle\| \bm{y} \right) = \int_{-\infty}^{\infty} f_{\bm{x}}(x)\log\left(\frac{f_{\bm{x}}(x)}{f_{\bm{y}}(y)}\right) $.}.
We demonstrate via a suite of numerical simulations in Section \ref{sec:sec7} that for a given $\epsilon>0$ and $\mathcal{F} \neq \emptyset$, there is a $T >0$ such that if robots use the proposed deployment strategy, $t>T$ implies $D_{\text{KL}}(\hat{b}_{t} || \hat{b}^{\star}) < \epsilon$.

%\Ram{How do you in practice solve this problem since we do not know $b_t^*$?}

\section{Probabilistic Range-limited Sensor Model}
\label{sec:sec3}
This sections present our combined sensor model. 
Each mobile robot is equipped with a \emph{range-limited sensor} that can measure quantitative information from afar and a \emph{radio} to communicate with other nodes to share its belief. 
%% MOTIVATION....
We  assume  that  a  sensor  can  correctly  measure  a  given target  only  if  the  sensor  can  detect  the  target \emph{a  priori}, and if  a  sensor  can  detect  a  target, a measurement of the target by the sensor may be corrupted. The combined sensor model joins the generic noisy sensor model with the binary detection model which generalizes existing methods \cite{viswanathan1997distributed,djuric2008target,anguelov2004detecting} to large-scale MSNs.
% The probabilistic range sensor model is, in fact, not new.
%Anguelov et al., \cite{anguelov2004detecting} apply the Expectation Maximization (EM) algorithm to cluster different types of objects from sequences of range data, where they adopted their such combined range sensor model. 
In fact, this combined sensor model has been experimentally validated during an object mapping and detection task using a laser scanner~\cite{anguelov2004detecting}.
%Their framework was validated experimentally on real-world datasets by showing the acquisition of accurate object maps and reliable detection.
%Although the sensor model was shown to follow the actual behavior of the laser scanner well \cite{anguelov2004detecting}, 
We postulate that this model is general enough to model other range-limited sensors as long as the sensor is capable of distinguishing the target from the environment and has uniform sensing range, i.e. 360-degree camera, wireless antenna, Gaussmeter, heat sensor, olfactory receptor, etc.
%to find more sensors with the characteristics, refer to \cite{akyildiz2002survey} and the reference therein).
%When detecting target, we assume that binary information (i.e., detection, no detection) is available. 
%During target detection, we assume that the sensor generates binary information corresponding to a detection.
%The binary detection model is used to capture one of the main characteristic of the atypical range sensors, that is, the chance of successful detection decreases monotonically by the distance between the sensor and the target \Ram{this sentence makes no sense}.
%The method can be also generalized if multi-level detection is possible.  
While performing the detection task, we assume that each sensor returns a $1$ if a target is detected or $0$ otherwise. 
The ability to detect a target for each $i^{\textup{th}}$ robot at time $t$ is a binary random variable $\bm{y}_{D,t}^i$ with a distribution that depends on the relative distance between the target and robot. 
%In other words, $\bm{y}_{D}^i=1$ is the event that the target is detected by the $i^{\textup{th}}$ sensor and $\bm{y}_{D}^i=0$ is the event that the target is not detected by the $i^{\textup{th}}$ sensor. 
This binary detection model, however, does not account for false positive or negatives.
For example, the probability of the event that all $m$ sensors with configuration $x_t$ fail to detect the target located at $q \in \mathcal{Q}$ is:
\begin{align}
&p\left(
\bm{y}_{D,t} = \bm{0} \mid {x}_{t},\bm{z}=(q,I)\right) = \prod_{i =1 }^{m}
p\left(
\bm{y}_{D,t}^i = 0 \mid
x_t,\bm{q}=q\right). \nonumber
\end{align}
For measuring a quantity of interest from a given environment, we consider a generic, noisy sensor model, where each sensor reports
binary output given a target state consisting of information and location.
The likelihood function at time $t$ is:
\begin{equation}
p(\bm{y}_t = \bm{1} \mid x_t, \bm{z} = (q,I)),
\label{lhd}
\end{equation}
which is the probability that $i$th robot measured the target 
with intensity value of $I$ at location $q$, i.e., positive measurement.
A general example of the likelihood function is a Gaussian, $\omega\mathcal{N}(I^{\star},\sigma_I^2)$ where $I^{\star}$ the ground truth intensity value at $q$, $\sigma_I^2$ is the variance of the intensity at the target located at $q$, and $\omega$ is a normalization constant. Note that since the observations made by $m$ robots are independent,
\[
p(\bm{y}_t = \bm{1} \mid x_t, \bm{z} = (q,I)) =
\prod_{i=1}^m
p(\bm{y}_t^i = 1 \mid x_t, \bm{z} = (q,I)),
\] 
or \eqref{lhd} can also be obtained via other distributed sensor fusion techniques (see e.g., \cite{stroupe2001distributed}). In our sensor model, we assume that at each $t$, the random vector $\bm{y}_t$ depends on $\bm{y}_{D,t}$, so that the conditional PDF can be computed as:
\begin{align*}
p(&\bm{y_t}=\bm{1} \mid x_{t},z) \\
=&p(\bm{y_t}=\bm{1} \mid x_{t},\bm{y}_{D,t} \neq \mathbf{0},z)\, p(\bm{y}_{D,t} \neq \mathbf{0} \mid x_{t},z) \nonumber\\
+&p(\bm{y_t}=\bm{1} \mid x_{t},\bm{y}_{D,t}=\mathbf{0},z)\, p(\bm{y}_{D,t} =\mathbf{0} \mid x_{t},z), \nonumber
\end{align*}
where $\bm{y}_{D,t} \neq \mathbf{0}$ means there is $j \in \lbrace 1,\dots,m \rbrace$ such that $\bm{y}_{D,t}^j = 1$ and $\bm{y}_{D,t} = \mathbf{0}$ means $\bm{y}_{D,t}^j = 0$ for all $j \in \lbrace 1,\dots,m \rbrace$. 
If the target cannot be detected, i.e., $\bm{y}_{D,t} =\bm{0}$, the measurement is taken as random and the likelihood function is modeled by uniform distribution, i.e.,
% \begin{align*}
% &p(\bm{y}_t = \bm{1} \mid z,x_{t},\bm{y}_{D,t}=\mathbf{0}) \\
$p(\bm{y}_t = \bm{1} \mid z,x_{t},\bm{y}_{D,t}=\mathbf{0})= \mathcal{U}(\mathcal{I})
$
% \end{align*}
supported on an interval $\mathcal{I}=[I_{\min},I_{\max}]$.
By the law of total probability,
%\begin{align*}
%&p_{\bm{y}_{D,t} \mid \bm{z},x_{t-1},u_t,\mathcal{F}}(\bm{y}_{D,t} =\mathbf{0}  \mid z,x_{t-1},u_t,\mathcal{F}) \\
%&= 1-p_{\bm{y}_{D,t} \mid \bm{z},x_{t-1},u_t,\mathcal{F}}(\bm{y}_{D,t} \neq \mathbf{0} \mid z,x_{t-1},u_t,\mathcal{F}).
%\end{align*}
\begin{align*}
&p(\bm{y}_t = \bm{1} \mid x_{t},z) =(1-
\underbrace{
p(\bm{y}_{D,t} = \mathbf{0} \mid x_{t},z)}_{(\star)}) \nonumber\\ 
& \times\underbrace{p(\bm{y}_t = \bm{1} \mid x_{t},\bm{y}_{D,t}\neq \mathbf{0},z)}_{\textup{likelihood of reliable measurements}}+\mathcal{U}(\mathcal{I})
\underbrace{p(\bm{y}_{D,t} = \mathbf{0} \mid z,x_{t})}_{(\star)}.
%\label{eq10}
\end{align*}
For the given target $z \in \mathcal{Z}$, by minimizing $(\star)$, the probability of missed-detection, one can ensure that the reliable measurements on the target state has been given more weight than the unreliable ones.

\section{Deployment Strategy}
\label{sec:sec4}
This section presents a class of deployment strategies for target detection capable of providing relative robustness.
At each time, $m$ robots move to new locations so as to minimize the missed-detection probability to promote the next observations. Since the set of robots with faulty sensors is unknown, we chose not to pose our problem to deal with the worst-case sensor failure scenarios which could be too conservative (e.g., the probability that multiple sensors fail at the same time is low). Instead, we adopt a provably optimal robot--target assignment method which can ensure that every target will be detected by at least one robot.
%By promoting the probability that at least one sensor can detect the target, one can ensure a worst-case guarantee.
%We consider a partitioned-based strategy where the workspace is partitioned into $m$ disjoint regions, and each robot is assigned to a region where it confines its detections 
%\Ram{just to be clear we are only going to consider the fully partitioned strategy in this paper?}.
%region is assigned to robots which will only attempt to detect targets in their regions, and will simply ignore target in other regions. 
This so called partitioned-based deployment is common to multi-robot coverage problems \cite{cortes_coverage_2004,hutchinson_robust_2012,park2014robust}. 
The most popular one is based on the Voronoi tessellations (see e.g., \cite{cortes_coverage_2004}, which we call a \emph{non-robust deployment}).
There are, in fact more general methods, which partition the workspace into $l$ regions and assign $k \in \lbrace 1,\dots,m \rbrace$ robots at each region (note that if $k=m$, the method becomes \emph{centralized}) \cite{hutchinson_robust_2012}. 
By doing so, one can ensure that each target has a chance to be detected by at least one of the $k$ sensors.
% such that the chance that target is missed detected by $m$ sensors decreases as we increase the value $k$. 
This approach, which we call the \emph{robust deployment}, can provide relative robustness by varying the value of $k$ from $2$ to $m$.
%Thus, if each sensor has an effective sensing range long enough to cover the whole workspace, utilizing all $m$ sensors to detect every target in the workspace becomes the most desirable strategy.

%\subsection{Robust detection}
\subsection{The Higher-Order Voronoi Partition for Robust Deployment}
% We will utilize the higher order Voronoi Tessellation to achieve robust robot--target assignment.
Recall that given $\mathcal{F}$, we want to ensure that at least one robot is detecting each target. One possible way of handling such robot--target assignment problem is the $k$-coverage method \cite{kumar2004k} which will guarantee that every target is covered by at least $k$ sensors. Another way is to use the higher order Voronoi partition, under which, for a given number of sensors (generators), exactly $k$ number of sensors are assigned to every region from the partition.
As long as $k \geq f+1$, if either of the two methods is used for the robot--target assignment, the constraint from \eqref{mainprob} will be satisfied.
Due to the bounded availability of sensor nodes, we will adopt the second approach in this study. 

Consider $m$ sensors and a workspace partition of $\mathcal{Q}$ into $l$ disjoint regions $W = (W^1,\dots,W^l)$, where $\cup_i W^i = \mathcal{Q}$, and $W^i \cap W^j = \emptyset$ for all $i \neq j$. 
Suppose the target location is a random variable $\bm{q}$ with a PDF, $f_{\bm{q}}:\mathcal{Q} \rightarrow \mathbb{R}_{\geq 0}$.
For a given target at $q \in \mathcal{Q}$, we define the probability that a sensor located at $x_i \in \mathcal{Q}$ can detect target, by using a real-valued function $h(\left\|q - x^i\right\|)$ as a probability measure%\footnote{For the numerical simulations purpose, we further assume that $h(\cdot)$ is continuously differentiable function non-increasing on its domain, and the image of $h$ must be in $[0,1]$ for it to be a probability measure.}
, which is assumed to decrease monotonically as a function of the distance between the target and the $i^{\textup{th}}$ sensor. 
Consider a bijection ${}^kG$ that maps a region to a set of $k$-points where the pre-superscript $k$ explicitly states that the region is mapped to exactly $k$ points.
%This association is used to decentralize the detection method.
Additionally we make the following definitions:
\begin{definition}[An Order$-k$ Voronoi Partition \cite{shamos1975closest}]
	Let $x$ be a set of $m$ distinct points in $\mathcal{Q}\subseteq \mathbb{R}^d$. 
	The \emph{order-$k$ Voronoi partition of $\mathcal{Q}$ based on $x$}\HJ{, namely ${}^kV$,} is the collection of regions that partitions $\mathcal{Q}$ where each region is associated with the $k$ nearest points in $x$.
	\label{orderk}
\end{definition}
Note that there is an $O(k^2n\log n)$ algorithm \cite{lee1982k} to construct the order-$k$ Voronoi diagram for a set of $n$ points in $\mathbb{R}^2$. We define another bijection ${}^kG^{\star}$ that maps a region to a set of $k$ \emph{nearest} points \HJ{(out of $x$) to the region.} 
%\Ram{what set of points?}. 
The total probability that all $m$ sensors fail to detect a target drawn by a distribution $f_{\bm{q}}$ from $\mathcal{Q}$ is:
\begin{equation}
\int_{\mathcal{Q}} p_{\bm{y}_{D} \mid x,\bm{q}
}\left(\bm{y}_D = \mathbf{0} \mid x,\bm{q}=q
\right)f_{\bm{q}}(q)\,{dq}.
\end{equation}
By substituting $\mathcal{Q}$ with the workspace partition $W$, and 
$p_{\bm{y}_{D} \mid x,\bm{q}
}\left(\bm{y}_D = \mathbf{0} \mid x,\bm{q}=q
\right)$
 with $h$, we have
\begin{align}
&H(x,W,{}^kG) := \nonumber \\
& \sum_{j = 1}^l \int_{W^j} \left( \prod_{x^i \in {}^kG(W^j)} \left(1- h \left(\left\|
q - x^i \right\|  \right) \right) \right)f_{\bm{q}}(q)\,dq
\label{cost2}
\end{align}
where we note again that the joint missed-detection events are conditionally independent, if conditioned on $x$.
In fact, the order-$k$ Voronoi tessellation is the optimal workspace partition which minimizes $H$ for each choice of $x$ and $k$:
\begin{theorem}[\cite{park2014robust}]
	For a given $x$ and $k$,
	$
	H(x,{}^{k}V,{}^kG^{\star}) \leq H(x,W,{}^{k}G)
	$
	for all $W$, ${}^{k}G$.
\end{theorem}
\noindent Note that the order-$k$ Voronoi partition $V_k$, along with the map ${}^kG^{\star}$ are uniquely determined given $x$, $f_{\bm{q}}$, and $\mathcal{Q}$. 

In addition, we introduce an additional constraint for the model, the \emph{effective sensing radius}, $r_{\text{eff}} >0$, to take into account the fact that each sensor has its own maximum sensing range. 
For a given $k$, $x$, $i$, and the target at $z = (q,I)$, our range-limited binary detection model in its final form becomes:
%{\small{
\begin{align}
%&\textit{Positive detection likelihood:} \nonumber \\
&p_{\bm{y}_{D}^i \mid \bm{q},x
}\left(\bm{y}_{D}^i = 1 \mid x,\bm{q}=q
\right) \nonumber\\
&=\begin{cases}
h\left(\left\|q-x^i \right\|\right), & \textup{if }q \in {}^kG^{\star}(x^i)
\cap \mathcal{B}(x^i,r_{\text{eff}})
,\nonumber \\
0, & \textup{otherwise},
\end{cases}\nonumber 
\end{align}
%\begin{align}
%&\textit{Negative detection likelihood:} \nonumber\\
%&p_{\bm{y}_{D,t}^i \mid \bm{z},x_t
%}\left(\bm{y}_t^i = 0 \mid x_t,z
%\right)\nonumber  \\
%&=\begin{cases}
%\left(1-h\left(\left\|q-x_t^i \right\|\right)\right) \times & \textup{if }q \in {}^kG_t^{\star}(x_t^i) \cap \mathcal{D}(x_t^i,r_{\text{eff}}), \\
%f_{\bm{y}_{I,t}^i \mid x_t,\bm{z}}(y_{I,t}^i \mid x_t,z), \\
%1,  & \textup{otherwise},
%\end{cases}
%\label{sensorf}
%\end{align}
%%}}
where $\mathcal{B}(p,r)$ is an open ball with radius $r$ centered at $p$.

\subsection{Gradient Algorithm for Deployment}
This section will present gradient descent-based deployment strategy.
Given the current configurations, robots solve decentralized counterpart of the original problem \eqref{mainprob}, move towards the solution, and the posterior belief is updated at robots' new  locations given the information collected from their sensors.
%Using (\ref{eq10}), we state the observation likelihood of positive detections as:
%\begin{align}
%&\int_{\mathcal{Q}}\int_{\mathcal{I}} f_{\bm{y}_{t} \mid \bm{z},x_t}(\bm{y}_{t}=\bm{1}\mid \bm{z}=(q,I),x_t)b_{t-1}(I \mid \bm{q} =q)\,dIdq \nonumber  \\
%&=\int_{\mathcal{Q}}\int_{\mathcal{I}} f_{\bm{y}_t \mid \bm{z},x_t,\bm{y}_{D,t},\mathcal{F}}(\bm{y}_t = \bm{1} \mid z,x_t,\bm{y}_{D,t}\neq \mathbf{0},\mathcal{F}) \nonumber  \nonumber \\
%&\,\,\,\,\,\,\,\times b_{t-1}(I \mid \bm{q} =q)f_{\bm{q}}(q)\,dIdq.\nonumber  \nonumber \\
%&-\int_{\mathcal{Q}}p_{\bm{y}_{D,t} \mid \bm{q},x_{t},\mathcal{F}}(\bm{y}_{D,t} = \mathbf{0} \mid \bm{q}=q,x_{t},\mathcal{F})  \nonumber \\
%&\,\,\,\,\,\,\,\times \int_{\mathcal{I}}f_{\bm{y}_t \mid \bm{z},x_{t},\bm{y}_{D,t},\mathcal{F}}(y_t \mid z,x_{t},\bm{y}_{D,t}\neq \mathbf{0},\mathcal{F})  \nonumber \\
%&\,\,\,\,\,\,\,\times b_{t-1}(I \mid \bm{q} =q)\,dIf_{\bm{q}}(q)\,dq.
%\label{cost0}
%\end{align}
%Recall that at each $t$, we are interested in finding the optimal control policy $u_t$ so as to drive robots to new configurations maximizing $\mathcal{L}^+(u_t)$ where $\mathcal{F} \subsetneq \lbrace 1,\dots,m \rbrace$. 
%For a given $q \in \mathcal{Q}$, let
%\begin{align*}
%\widetilde{f}_{\bm{q}}(q) &:= \int_{\mathcal{I}}f_{\bm{y}_t \mid \bm{z},x_{t-1},u_t,\bm{y}_{D,t},\mathcal{F}}(y_t \mid z,x_{t-1},u_t,\bm{y}_{D,t}\neq \mathbf{0},\mathcal{F})) \\
%&\,\,\,\,\,\,\,\times b_{t-1}(I \mid \bm{q} =q)\,dI\,f_{\bm{q}}(q).
%\end{align*}
 By using $f_{\bm{q}}$, and \eqref{cost2}, for a given $x_{t}$, we want to obtain the next way-point $x_t^{\star}$ by
\begin{align}
&x_{t}^{\star} \gets \arg\min_{x_{t}}\Biggl\{\mathcal{L}(x_{t}):=\Biggr.
%&:=\int_{\mathcal{Q}}p_{\bm{y}_{D,t} \mid \bm{q},x_t}(\bm{y}_{D,t} = \mathbf{0} \mid \bm{q}=q,x_t)\widetilde{f}_{\bm{q}}(q)\,dq \nonumber \\
%&= \int_{\mathcal{Q}} \prod_{i =1}^m 
%p_{\bm{y}_{D,t}^i \mid
%	{x}_{t},\bm{q}}\left(
%\bm{y}_{D,t}^i = 0 \mid
%{x}_{t},\bm{q}=q\right) \widetilde{f}_{\bm{q}}(q)\,dq \nonumber\\
%&=  
%\int_{\mathcal{Q}} \prod_{i =1}^m \left( 1-h\left(\left\|q-x_t^i \right\|\right) \right) \widetilde{f}_{\bm{q}}(q)\,dq 
%&\int_{\mathcal{Q}}
%p_{\bm{y}_{B,t-1} \mid
%	{x}_{t},\bm{q}}\left(
%\bm{y}_{B,t-1} = \bm{0} \mid
%{x}_{t},q\right)
%\widetilde{b}_{t-1}(q)
%\,dq 
\nonumber \\
\Biggl.& \sum_{j = 1}^l \int_{W_t^j} \prod_{x_{t}^i \in {}^kG^{\star}(W_t^j)} \left(1- h \left(\left\|
q - x_t^i \right\|  \right) \right){f}_{\bm{q}}(q)\,dq \Biggr\}.
\label{cost1}
\end{align}
where $\mathcal{L}(x_t)$ takes the identical form as \eqref{cost2} by adding subscript $t$ to $x^i$s and $W^i$s.
% \eqref{cost1} takes the identical form as $H(x,W,G^k)$ which was previously defined in \eqref{cost2}.
%We note that for a given $k$, by substituting $\widetilde{f}_{\bm{q}}$ with $\phi$ and $x$ with $x_t$, \eqref{cost1} takes the identical form as $H(x,W,G^k)$ which was previously defined in \eqref{cost2}.
%The value of $\overline{\mathcal{L}}_k^+(u_t)$ depends on $u_t$, the maximum number of faulty robots. In essence, the term $\overline{\mathcal{L}}_k^+(u_t)$ accounts for an addition of random effort due to the likelihood of target missed-detection when $k$ robots are assigned to each target.
%Thus, our problem become
%\begin{equation}
%{u}_t^{\star} = \argmin \overline{\mathcal{L}}_k^+(u_t),
%\label{mmle}
%\end{equation}
%subject to the vehicle dynamics (\ref{eq:constraint}). 
%If $\widetilde{f}_{\bm{q}}$ takes uniform density on $\mathcal{Q}$, then the physical meaning of the problem (\ref{mmle}) is to find control law minimizing the probability that all robots fail to detect target over $\mathcal{Q}$ when $k$ robots are assigned to each region, i.e.,  maximizing the probability that at least one robot can detect target over $\mathcal{Q}$. 
%which illustrates the explicit dependence on the previous belief $\tilde{b}_{t-1}$ and previous observations $y_{t-1}$. 
%Due to the way we posed the problem, if $k=n$, the solution to \eqref{mmle} guarantees maximum observation likelihood for up to $k-1=n-1$ faults, whereas if $k=1$, the solution does not provide any robustness guarantee even for a single faulty robot.
%The problem is NP-Hard for $k=1$ and so on.
If $h$ is differentiable, our deployment strategy can use the gradient $\nabla  \mathcal{L}(x_t)=\left[\frac{\partial\mathcal{L}(x_t)}{\partial x_1^t},\dots,\frac{\partial\mathcal{L}(x_t)}{\partial x_m^t}\right]$ where for each $i$,

\begin{align*}
\frac{\partial\mathcal{L}(x_t)}{\partial x_t^i} 
&= -\sum_{j \in \lbrace 1,\dots,l \rbrace: \atop W_t^j \in {}^kG^{\star-1}(x_t^i)} \int_{W_t^j}
\frac{\partial h(\left\| q-x_t^i \right\|)}{\partial x_t^i} \\
&\,\,\,\,\,\,\, \times \prod_{l \in \lbrace 1,\dots,m \rbrace: \atop x_{t}^l \in {}^kG^{\star}(W_t^j),l\neq i} \left(1- h \left(\left\|
q - x_t^i \right\|  \right) \right){f}_{\bm{q}}(q)\,dq,
\end{align*}
to find the desirable way-points of the robots as described in Algorithm \ref{alg1}. For each $t$, Algorithm \ref{alg1} uses coordinate gradient descent in cyclic fashion\footnote{A general version of Algorithm \ref{alg1}, which uses block coordinate descent, has been shown to be convergent using the Invariance Principle \cite{park2014robust}.} to converge to a sub-optimal solution, namely, $\hat{x}_t^{\star}$.
{\tiny{
		\begin{algorithm}
			\DontPrintSemicolon
			\KwIn{$\mathcal{L}_k,\,\hat{x}_{t},\,\epsilon >0 $}
			\KwOut{$\hat{x}_t^{\star}$}
			$k \gets 0$, $\Delta \gets \epsilon$\\
			\While{$\Delta > \epsilon$}{
			\ForEach{$i \in \lbrace 1,\dots,m \rbrace $}{
				$x_{t,k+1}^i \gets x_{t,k}^i- \alpha_{t,k}^i 
				\nabla_{i} \mathcal{L}_k(x_{t,k}) $ \\
				\tcp{$\alpha_{t,k}^i$ is a step-size obtained using a line search method}
	}
				$\Delta \gets
 \mathcal{L}_k(x_{t,k})  - 
 \mathcal{L}_k(x_{t,k+1}) $	\\
$k \gets k +1$
}
$\hat{x}_t^{\star} \gets x_{t,k}$,
\Return $\hat{x}_t^{\star}$
			\caption{Gradient Algorithm}\label{alg1}
		\end{algorithm}
}}

\section{Implementation: Environmental Mapping}
\label{sec:sec6}
In this section, we first introduce Bayesian filtering equations for our particular target distribution, and then present a particle filter to reduce the complexity of \HJ{the map construction process.} 
%\Ram{how is the target map evolving? I thought it was fixed...?I think you mean the reconstruction is being updated as observations appear...}. 

\subsection{Recursive Bayesian Filter}
We present a brief overview of the Bayesian filter, and the derivation of the filtering equations for our primary goal: environmental mapping by $m$ robots. 
Recall that $b_t(z)$ represent a \emph{belief} on target state---the posterior probability distribution of the target state described by a random vector $\bm{z} \in \mathcal{Z}$---at time $t \in \mathbb{Z}_{\geq 0}$. 
%Recall that the belief about a given target state $z$ at time $t \in \mathbb{N}$, i.e., an environmental map at time $t$, was denoted by $b_t(z)$. 
In a similar manner, the belief of target information state $I$ given the target located at $q$ is
\begin{align}
&b_t(I \mid \bm{q} = q) = f_{\bm{I}\mid b_0,{x}_{0:t}\bm{y}_{1:t},\bm{q}}\left(I \mid b_0,{x}_{0:t},y_{1:t},\bm{q}=q\right)
\label{eqb}
\end{align}
where we denote the initial belief on target state by $b_0$.
%Note that $b_t(I \mid \bm{q} = q)$ depends on the initial belief $b_0$, the previous robot trajectories up until time $t$ where $t\in \mathbb{Z}_{\geq 0}$, and observations up to this point, $\bm{y}_{1:t}$. 
The belief on the complete target state $\bm{z}$ is:
\begin{align}
b_t(z)&=
f_{\bm{z} \mid b_0,{x}_{0:t},\bm{y}_{1:t}}
\left(
z \mid b_0,
{x}_{0:t},y_{1:t}
\right) =
b_t(I\mid \bm{q}=q)f_{\bm{q}}(q). 
\label{eq0}
\end{align}
%\Ram{I don't understand the marginal in the first line of the previous equation. Are you saying that we know $b_0$ or $b_0(z)$? Those are two distinct objects.}
%If there is no prior knowledge of the target information at the initial time, one can choose the prior distribution as the \emph{uniform} density. 
%Recall that the generic Bayes' Theorem states,
%$
%P(B\mid A) = \frac{P(A \mid B)P(B)}{P(A)}
%$
%where $A$, $B$ are events, $P(B\mid A)$ is the posterior probability distribution, $P(A \mid B)$ is the likelihood function, $P(B)$ is the prior probability distribution, and $P(A)$ is the marginal probability of the event $A$. Let the event $B$ represent the prior belief on target information state $b_{t-1}(I\mid \bm{q}=q)$, and $A$ represent the sensor measurements at time $t$, $\bm{y}_t$, then 
In our problem, the observation $\bm{y}_{t}$ is conditionally independent of $b_0$, $\bm{y}_{1:t-1}$, and $x_{0:t-2}$ when it is conditioned on $\bm{z}$ and $x_t$. Applying \emph{Bayes' Theorem}, (\ref{eqb}) becomes:
%{\small{
%		\begin{align*}
%		&b_{t}(I \mid \bm{q} =q)=
%		\frac{
%			f_{\bm{y}_{t} \mid
%				\bm{z},
%				{x}_{t}
%			}\left(
%			y_{t} \mid
%			\bm{z}=(I,q),
%			{x}_{t}
%			\right)
%			b_{t-1}(I\mid \bm{q} = q)
%		}
%		{
%			f_{\bm{y}_{t} \mid
%				\bm{q},
%				{x}_{t}	
%			}\left(
%			\bm{y}_{t} = \bm{1} \mid
%			\bm{q}=q,
%			{x}_{t}
%			\right)
%		}
%		\end{align*}}}where $t \in \mathbb{N}$.
%%\Ram{I still do not understand this previous equation. Here are my questions: 1) is it $b_0(z)$ or $b_0$ 2) is it $x_{0:t}$ or $x_{0:t-1}$? }.
%%\Ram{again is this supposed to be $b_0(z)$ or $b_0$?}. 
%One can simplify the likelihood function in the target information map by using this observation, which yields:
\begin{align}
&b_{t}(I \mid \bm{q} = q)= \nonumber \\
&\eta_t\,
f_{\bm{y}_{t} \mid
	\bm{z},{x}_{t}}\left(
\bm{y}_{t} = \bm{1} \mid
\bm{z}=(I,q),{x}_{t}\right)
b_{t-1}(I \mid \bm{q} = q)
\label{eq1}
\end{align}
where 
$
\eta_{t} \coloneqq \left(
f_{\bm{y}_{t} \mid
	\bm{q},
	b_0,
	{x}_{t}	
}\left(
\bm{y}_{t}=\bm{1} \mid
\bm{q}=q,b_0,
{x}_{t}
\right)
\right)^{-1}
%\label{eq2}
$ is a \emph{normalization constant}. 
%
% which usually cannot be directly computed, but can be obtained by utilizing the total law of probability:
%\begin{align*}
%\eta_{t} =
%&\bigg(\int_{\mathcal{I}}
%f_{\bm{y}_{t} \mid
%	\bm{z},
%	{x}_{t}	
%}\left(
%y_{t} \mid
%\bm{z}=(I,q),
%{x}_{t}
%\right) b_{t-1}(I\mid \bm{q} = q)
%\,dI\bigg)^{-1}
%\end{align*}
By joining the \eqref{eq0} and \eqref{eq1}, one can obtain a simplified form of the filtering equation:
\begin{align*}
b_t(z) &=
\eta_t\,
f_{\bm{y}_{t} \mid
	\bm{z},{x}_{t}}\left(
\bm{y}_{t} = \bm{1} \mid
z,{x}_{t}\right)
b_{t-1}(z) \nonumber\\
&=\left(\prod_{i=1}^t
\eta_i
f_{\bm{y}_{i} \mid
	\bm{z},{x}_{i}}\left(
\bm{y_{i}}=\bm{1} \mid
z,{x}_{i}\right) \right)
b_0(z).
\end{align*}
%We assume that $m$ robots share their beliefs.
% \Ram{so you are assuming full information sharing?} 

\subsection{Belief Approximation via SIR Particle Filter}
For our numerical simulations, we consider a low discrepancy sampling method, namely, \emph{Halton-Hammersley sequence}, to sample continuously distributed targets in $\mathcal{Z}$. 
This approach has been used for sampling-based algorithms for robot motion planning \cite{lavalle2006planning}.
We consider Sequential Importance Resampling (SIR) for the particle filtering process.
For a given distribution on target locations, $f_{\bm{q}}(q)$, at each time $t$, based on the observations, the locations belief hypothesis is populated for $N_1$ samples initially generated with Halton-Hammersley sequence. In a similar manner, for each sample $q^{i}$
the information belief hypothesis is populated for $N_2$ samples from $\mathcal{I}$ initially generated by the Halton-Hammersley sequence.
%\begin{equation*}
$q^{1},\dots,q^{N_1}$
%\end{equation*}
%where $w_T^{(i)} = 1/N_1$ for all $i$.
Thus, for each $i \in \lbrace 1,\dots,N_1 \rbrace$, $j \in \lbrace 1,\dots,N_2 \rbrace$,
%\begin{equation}
%\widetilde{w}_t^{i} =
%p_{\bm{y}_{D,t} \mid
%	\hat{x}_{t},\bm{z}}\left(
%y_{D,t} \mid
%\hat{x}_{t},\bm{z}=(q_t^i,I)\right) 
%\end{equation}

\begin{equation*}
\widetilde{w}_t^{ij} \propto f_{\bm{y}_t \mid \bm{z}_t,x_t}(\bm{y}_t = \bm{1} \mid \hat{x}_t,\, \bm{z}=(q^i,I^{ij})).
%p_{\bm{y}_{D,t} \mid
%	\hat{x}_{t},\bm{z}}\left(
%y_{D,t} \mid
%\hat{x}_{t},\bm{z}=(q_t^i,I)\right) 
\end{equation*}
 If we let $z_t^{ij} \coloneqq (q^{i},I_t^{ij})$, then the collection of $N\coloneqq N_1\times N_2$ tuples---where each tuple is a particle-weight pair---is:
\begin{align*}
\lbrace
\lbrace
(
z^{i1},\widetilde{w}_{t}^{i1}),\dots,(
z^{iN_2},\widetilde{w}_{t}^{iN_2}) \rbrace_{i\in \lbrace 1,\dots,N_1\rbrace}\rbrace
%&\left\{\left(
%z^{21},\widetilde{w}_{t}^{21}\right),\dots,\left(
%z^{2N_2},\widetilde{w}_{t}^{2N_2}\right) \right\},\dots, \\
%&\left. \left\{
%\left(
%z^{N_11},\widetilde{w}_{t}^{N_11}\right),\dots,\left(
%z^{N_1N_2},\widetilde{w}_{t}^{N_1N_2}\right) \right\} \right\}
\end{align*}
where for each $t$ and $i = 1,\dots,N_1$, $\sum_{j=1}^{N_2} \widetilde{w}_{t}^{ij} = 1$.
%\begin{equation}
%\widetilde{w}_{I,t}^{ij} \propto
%p_{\bm{y}_{I,t} \mid
%	\hat{x}_{t},\bm{z}}\left(
%y_{I,t} \mid
%\hat{x}_{t},\bm{z}=(q^i,I^{ij})\right)
%\end{equation}
%for all $i \in \lbrace 1,\dots,N_1 \rbrace$.
%\begin{equation}
%\left\{
%\left\{
%(z^{ij},\widetilde{w}_t^{ij})
%\right\}_{j=1}^{N_2}
%\right\}_{i=1}^{N_1}
%= \left\{
%z^k,\widetilde{w}_t^k
%\right\}_{k=1}^N
%\end{equation}
After resampling and normalizing, the approximate belief becomes
\begin{equation*}
\hat{b}_t(z) = \sum_{k=1}^{N} 
w_t^{k} \delta(z - z^{k})
\end{equation*}
% which is a form of discrete random measure 
where the $w_t^1,\dots,w_t^N$ are resampled, normalized weight such that $\sum_{k=1}^{N} w_t^{k} = 1$, and $\delta(z - z^{k})$ is Dirac-delta function evaluated at $z^{k}$. 
%We note that resampling is only taken on the target information state, namely, $\bm{I}_t$.
The whole filtering process is depicted in Algorithm \ref{costalgorithm}.
Note that as discussed in previous studies \cite{crisan2002survey}, our particle filter uses a standard re-sampling scheme to ensure the convergence of the mean square error toward zero with a convergence rate of $1/N_2$ for all $q \in \mathcal{Q}$. 
{\tiny{
\begin{algorithm}
	\DontPrintSemicolon
	\KwIn{$\hat{b}_{t-1}= \lbrace z^l,\,w_{t-1}^l \rbrace_{l=1}^N 
		,y_{t},\hat{x}_{t},\,I_{\textup{range}}$}
	\KwOut{$\hat{b}_t$}
%	\tcp{Propogate motion model; see Algorithm \ref{alg1}}
%	$\hat{u}_{t}\gets$ \text{MMLE}($y_{D,t}$, $\hat{x}_{t-1})$ \\
	\tcp{SIR Particle Filter}
	\tcp{1) Update using the observation model}
		\ForEach{$i \in \lbrace 1,\dots,N_1\rbrace$}{
%		{$\widetilde{w}_{D,t}^{i} \gets
%		p_{\bm{y}_{D,t} \mid
%			\hat{x}_{t},\bm{q}}\left(
%		y_{D,t} \mid
%		\hat{x}_{t},\bm{q}=q_t^i\right)$ \\
		\ForEach{$j \in \lbrace 1,\dots,N_2 \rbrace$}{
			$\widetilde{w}_{t}^{ij} \gets 
			p_{\bm{y}_{D,t} \mid \bm{q},\hat{x}_{t}}(\bm{y}_{D,t} = \mathbf{0} \mid \bm{q}=q^i,\hat{x}_t)  
			( I_{\textup{range}}^{-1} -  w_{t-1}^{ij} f_{\bm{y}_t \mid \bm{z},\hat{x}_{t},\bm{y}_{D,t}}(\bm{y}_t = \bm{1} \mid \bm{z} = z_t^{ij},\hat{x}_{t},\bm{y}_{D,t}\neq \mathbf{0}))
			+ w_{t-1}^{ij}f_{\bm{y}_t \mid \bm{z},\hat{x}_{t},\bm{y}_{D,t}}(\bm{y}_t = \bm{1} \mid z,\hat{x}_{t},\bm{y}_{D,t}\neq \mathbf{0})
			 $
	}
}
	\tcp{2) Resample and Normalize}
	$\lbrace w_{t}^l \rbrace_{l=1}^N \gets$ \text{Resample}$(\lbrace \widetilde{w}_{t}^l \rbrace_{l=1}^N,\,\lbrace w_{t-1}^l \rbrace_{l=1}^N)$ \\			
	\Return{$\hat{b}_t \gets \lbrace z^l,w_t^l \rbrace_{l=1}^N $}
	\SetKwBlock{Begin}{function}{end function} \\
		\tcp{Low Variance Resampling \cite{choset2005principles}}
	\Begin($\text{Resample} {(} \lbrace \widetilde{w}_{t}^l \rbrace_{l=1}^N,\,\lbrace w_{t-1}^l \rbrace_{l=1}^N {)}$)
	{

%		$S_{i,c_{i}} = \left[ \right]$\;
		\ForAll{$i \in \lbrace 1,\dots,N \rbrace$}
		{
			$
			\overline{w}_{t}^{i} \gets \frac{
			\widetilde{w}_{t}^{i} \cdot w_{t-1}^{i}}{\sum_{i = 1}^{N} \widetilde{w}_{t}^{i} \cdot w_{t-1}^{i}}$ 
		} 
	
	\ForEach{$i \in \lbrace 1,\dots,N_1 \rbrace$}{
		$\delta \gets \textup{rand}((0;N_2^{-1}))$  \\
		$cdf \gets 0,$ $k \gets 0,$ $c_j\gets []$ for all $j$\\
		\For{$j=0,\,j<N_2$}{
		$u \gets \delta + j \cdot{N_2}^{-1}$ \\
		\While{$u > \text{cdf}$}{
		$k \gets k+1$	\\
		$cdf \gets cdf + \overline{w}_{t}^{ik}$
	}
		$c_{j+1} \gets k$		 	
	}
	\For{$j=1;\,j \leq  N_2$}{
		$w_{t}^{ij} \gets \frac{c_j}{N_2}$
		}
}
\Return{
$\hat{b}_t=\lbrace 
z^l,\,w_{t}^l
\rbrace_{l=1}^{N}$
}
	}
	\caption{Filtering Algorithm}\label{costalgorithm}
\end{algorithm}
}}
\section{Numerical Simulations}
\label{sec:sec7}
This section presents a suite of numerical simulations to validate both our sensor model and deployment strategy under sensor failures.
%\textit{Gaussian PDFs for the Observation Likelihood:}
%For the simulation, we consider Gaussian kernels for the probability distributions of both the perception model, and the detection model.
%First, consider the conditional probability distribution for detection likelihood, positive and negative functions respectively
%\begin{align*}
%&p_{
%	\bm{y}_{D,t}^i\mid
%	x_t,\bm{q}
%}(\bm{y}_{D,t}^i=1\mid x_t,\bm{q}=q)
%=\eta_{D}\mathcal{N}(q,x_t^i,\,\Sigma_B) \\
%& = \eta_{D}\frac{1}{2u\left|\Sigma_B\right|}
%\exp\left(
%\frac{-(q-x_t^i)^{\top}\Sigma_B^{-1}(q-x_t^i)}{2}
%\right),
%\end{align*}
%and 
%\[
%p_{
%	\bm{y}_{D,t}^i\mid
%	x_t,\bm{q}
%}(\bm{y}_{D,t}^i=0\mid x_t,\bm{q}=q)
%=1-\eta_{D}\mathcal{N}(q,x_t^i,\,\Sigma_B)
%\]
%where $\mathcal{N}(q,x_t^i,\Sigma_B)$ is multivariate Gaussian with mean $x_t^i$ and covariance matrix $\Sigma_B$, and $\eta_{D}$ is a constant.
%Assume that the noisy sensor model is also a multi-variate Gaussian with mean $y_{I,t}^i$ and covariance matrix $\Sigma_I$.
%\[
%f_{
%	\bm{y}_{I,t}^i\mid
%	x_t,\bm{I}
%}(y_{I,t}^i\mid x_t,\bm{I}=I)
%=
%\eta_I \mathcal{N}(I,y_{I,t}^i,\Sigma_I)
%\]
%where $\eta_I$ is normalization constant. The total observation likelihood is given by
%\begin{align*}
%&f_{\bm{y}_t^i\mid x_t,\bm{z}}
%\left(
%y_{D,t}^i,y_{I,t}^i \mid x_t,\bm{z}=(q,I)
%\right) \\
%&=\begin{cases}
%\eta_B\eta_I(1-\mathcal{N}(q,x_t^i,\,\Sigma_B))\mathcal{N}(I,y_{I,t}^i,\Sigma_I)
%, &\textup{if }y_{D,t}^i = 0 \\
%\eta_B\eta_I\mathcal{N}(q,x_t^i,\,\Sigma_B)
%\mathcal{N}(I,y_{I,t}^i,\Sigma_I)
%, &\textup{if }y_{D,t}^i = 1
%\end{cases}
%\end{align*}

\noindent \textit{Simulation Settings:}
Let $\mathcal{Q}$ be a rectangular space $[42.00,41.51]\times [-73.49,-72.83]$ in $\mathbb{R}^2$ which corresponds to a mountainous region in Connecticut, U.S.A, where each coordinate corresponds to latitude and longitude, respectively. 
We let $\mathcal{I} = [-1000,\,4000]$ be a range of elevations in feet.
%and $r_{\text{eff}} =2.89$ miles. 
Targets are uniformly distributed over $\mathcal{Q}$, and the ground truth target information over $\mathcal{Q}$ is depicted in Fig. \ref{fig:fig3} (right). 
% The information vector ranges from $-1000$ to $4000$, i.e., $\mathcal{I} = [-1000,\,4000]$.
The robots have no prior knowledge of the target information.
A number of particles used for the SIR filter is $N= N_1\times N_2 = 5000 \times 100$. 
We consider Gaussian distribution to for both the perception and the detection model. 
Each sensor's measurement noise covariance matrix is $\Sigma_I = 0.5\mathbf{I}$, and the binary detector's noise covariance matrix is $\Sigma_B = 0.04\mathbf{I}$ where $\mathbf{I} \in \mathbb{R}^{d\times d}$ is an identity matrix.
In our simulation, we compare the three methods summarized in Table \ref{table2}. 
% Note that the three methods presented here are not exactly same as those found from the reference therein, nevertheless, we postulate that the results will be comparable due to the similarities of the ideas.
\begin{table}[]
	\centering
	\caption{Summary of deployment methods considered in current section.}
				\label{table2}	
	{\scriptsize
		\resizebox{\linewidth}{!}{%	
			\begin{tabular}{lll}
				\hline
				algorithm type:           & gradient computation & related studies \\
				\hline 
				non-robust & fully decentralized & \cite{cortez2011information,bandyopadhyay2003energy,patten2013large} \\ 
%				\hline 
				robust ($k=2$) & decentralized & current paper \\
%				\hline
				max. information gain & centralized & \cite{connor2016airborne,schwager2017multi,pahlajani2014networked,julian2012distributed,lynch2008decentralized} \\
				\hline
	\end{tabular}}}
	\vspace*{-0.25cm}
\end{table}
%\begin{figure}
%	\centering
%	\includegraphics[width=1.6in]{figure/gtruth2d2}
%	\caption{Elevation map of some region in Connecticut (the ground truth).} 
%	\label{fig:fig1}
%\end{figure}

%\begin{figure}
%	\centering
%	\includegraphics[width=1.6in]{figure/init}
%	\caption{Expected target intensity for an arbitrary distribution (the ground truth)} 
%	\label{fig:fig1}
%\end{figure}
\begin{figure}
		\centering
		\includegraphics[width=3.35in]{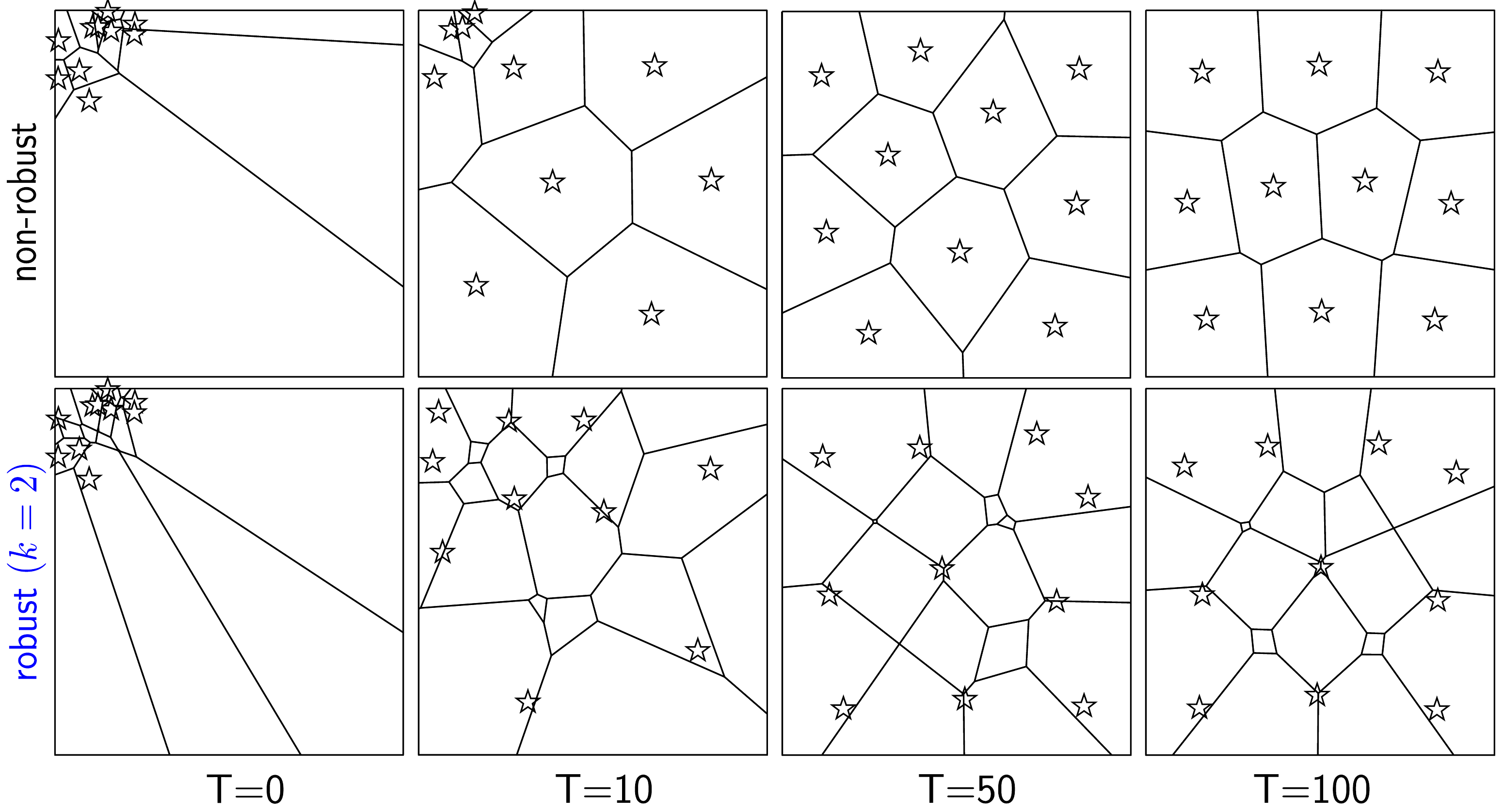}
% 	\begin{subfigure}[b]{0.23\textwidth}
% 		\centering				
%   	\includegraphics[width=1.4in]{figure/new_path1}
% 	\caption{}
% \end{subfigure}  
% 	\begin{subfigure}[b]{0.23\textwidth}
% 	\centering		
%   	  	\includegraphics[width=1.4in]{figure/new_path2}
% 	\caption{}
% \end{subfigure}    	  	
%\includegraphics[width=1.57in]{figure/final_config_order1_fault}
	%	\includegraphics[width=3.4in]{figure/init_10_deploy_cmd2}
	\caption{An illustration of the convergence of two different deployment strategies, top: non-robust, bottom: robust ($k=2$) (stars: positions of robots, polygons: partition).}
	\label{fig:fig2}
\end{figure}
\begin{figure}
    \vspace*{-0.15in}
	\centering
	\begin{subfigure}[b]{0.23\textwidth}
		\centering
	\includegraphics[width=1.8in,keepaspectratio=true]{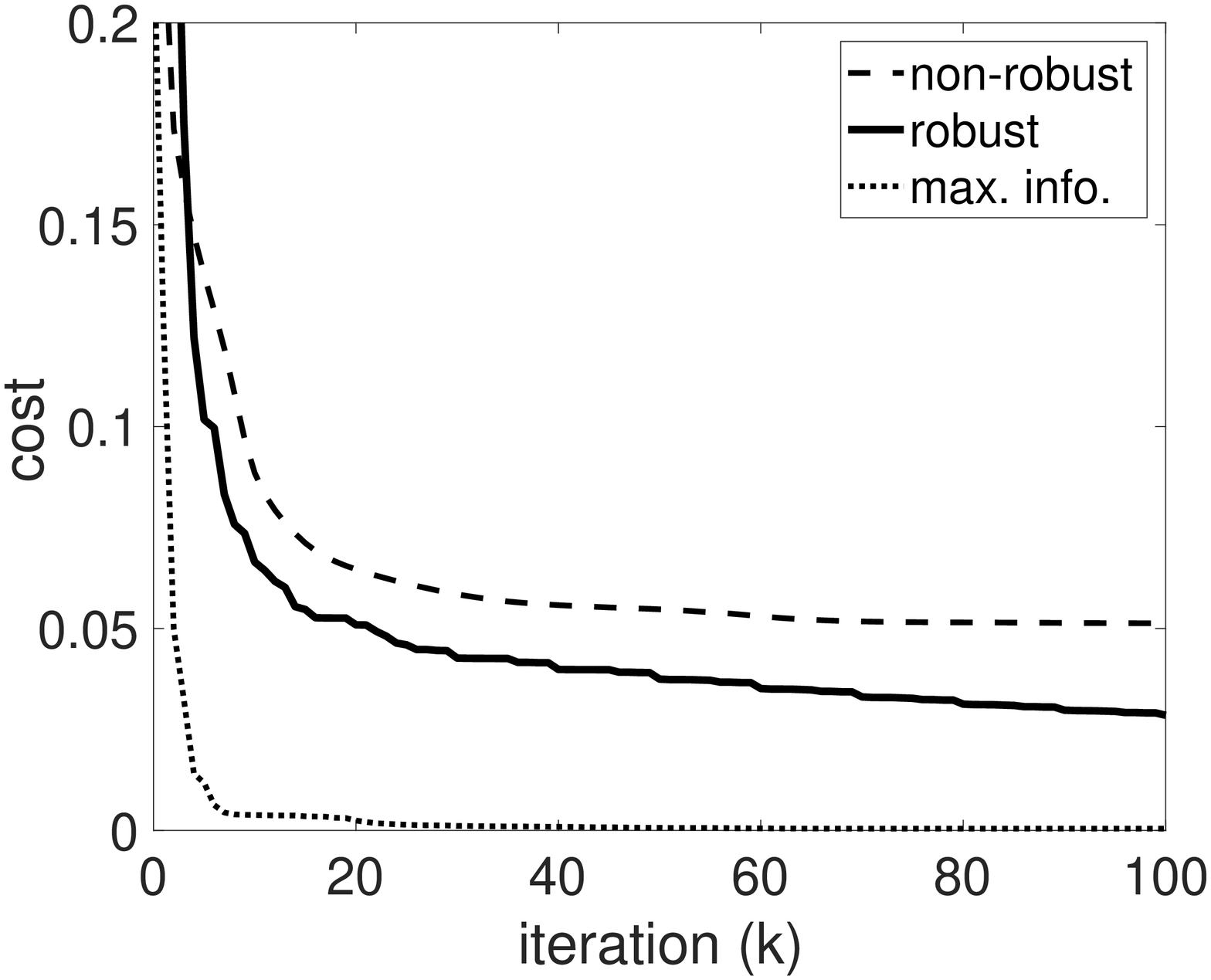}
	\caption{}
	\end{subfigure}
	\begin{subfigure}[b]{0.23\textwidth}
		\centering
	\includegraphics[width=1.8in,keepaspectratio=true]{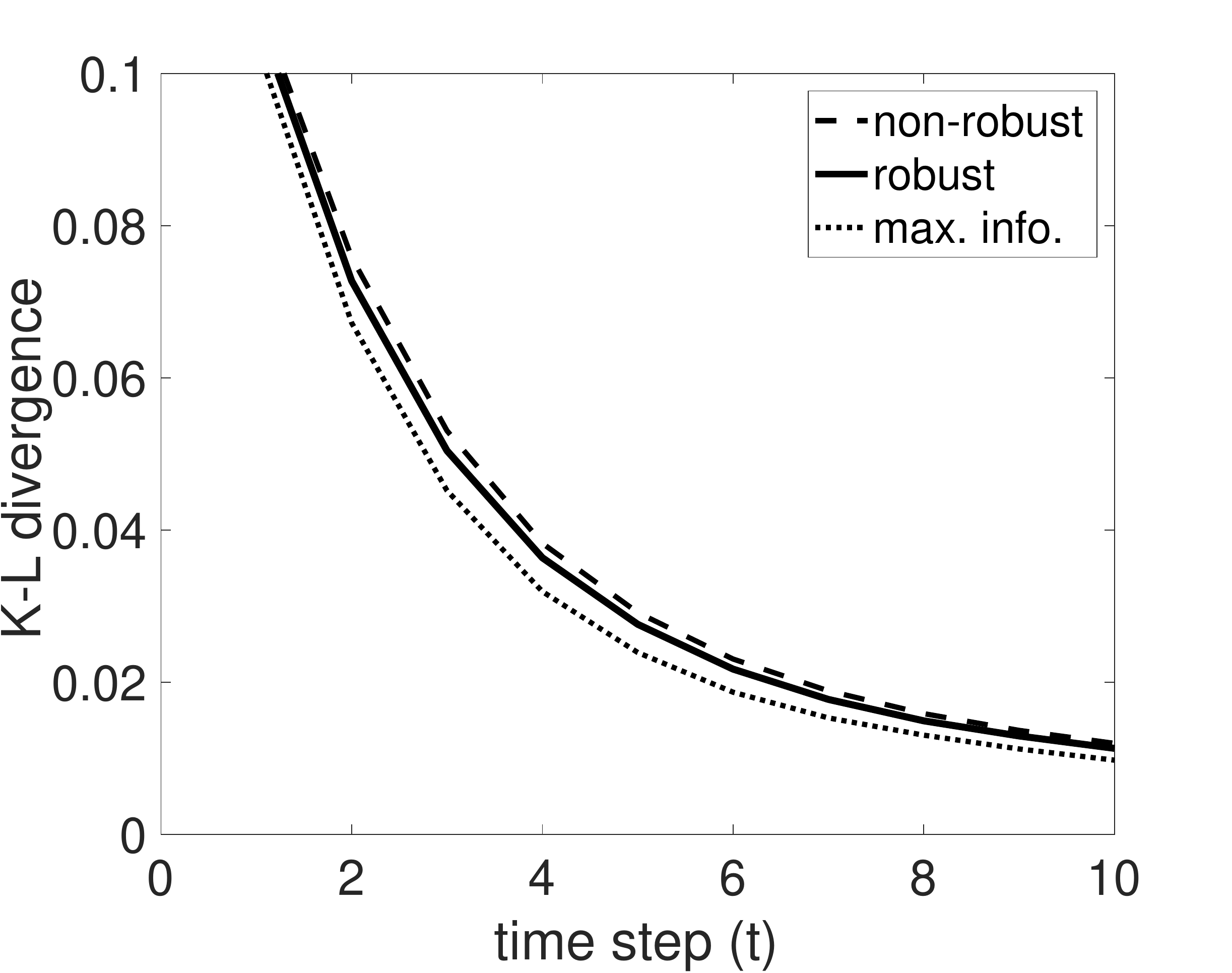}
	\caption{}
	\end{subfigure}
	\caption{(a) convergence test for one-time deployment with different methods, (b) comparison of K-L divergence values between different methods.}
	\label{fig:fig0}
\end{figure}
%\begin{figure}
%	\centering
%	\includegraphics[width=1.1in]{figure/order1deploy3d}
%	\includegraphics[width=1.1in]{figure/order2deploy3d}
%%	\includegraphics[width=3.4in]{figure/init_10_deploy_cmd2}
%	\caption{One-time deployment with $k=1$, (a) cost change during Algorithm 1 ($0^{\textup{th}}-1^{\textup{st}}$ step) , (b) optimal path respecting unicycle kinematic constraint (circles: initial positions, stars: target positions, polygons: partition)}
%	\label{fig:fig2}
%\end{figure}

\noindent \textit{Convergence of Our Deployment Strategy:}
First, the behavior of the deployment strategy is discussed. 
Given a uniform initial prior belief and an initial configuration at $t=0$ (Fig. \ref{fig:fig2} (top-left)), three algorithms, summarized in Table \ref{table2}, were tested.
Fig. \ref{fig:fig2} shows positions of robots after the $T$ number of iterations with (a) the non-robust method  and (b) robust method ($k=2$). 
Fig. \ref{fig:fig0}(a) compares the convergence speeds between the three methods.
The cost on the $y$-axis corresponds to the probability of missed detection. 
%Notice that the robust method has lower cost than the non-robust method even though it relies on a decentralized deployment strategy.
Notice that the maximum information algorithm has a lower cost, but it relies on a centralized scheme that is typically impossible to realize in practice.

\noindent \textit{Environmental Mapping/Filtering Performance Without Sensor Failure:}
Next, we present the evolution of the belief to build an estimate of the elevation map after the deployment strategy has been completed. 
%given the uniform, initial map with successive positive observations, each followed by the filtering process.
%and the gradient descent strategy. 
%Fig. \ref{fig:fig3} illustrates the map building process, given limited effective sensing range value ($r_{\textup{eff}}=0.3$, the value was chosen empirically relative to the workspace size) when using our robust method. Others show similar performance, and are omitted due to the lack of space.
%
%
%The results in Fig. \ref{fig:fig3} clearly show that under \emph{limited} sensing range, the non-coordinated strategy ($k=1$) yields relatively better mapping results than coordinated strategies ($k=2,\,10$) compared to the ground truth map shown in Fig \ref{fig:fig1}.
Fig. \ref{fig:fig0}(b) compares the K-L divergence values between the different strategies during the filtering process. 
While the maximum information gain approach shows the best result, the robust deployment strategy has competitive mapping performance relative to the ground truth despite being a decentralized approach.
%The occurrence of sudden jumps (between the time step 0 and 1) in the K-L divergence values observed in Fig \ref{fig:fig5}(a) demonstrates the cases when the initial uniform density happened to a better `\emph{guess}' than the crude belief obtained after a single propagation of the filtering process.
%\begin{figure}
%	\centering
%	\includegraphics[width=2.9in]{figure/cost_comp0}
%	\caption{Comparison of K-L divergence from the actual distribution between different methods during belief propagation.}
%	\label{fig:fig5}
%\end{figure}
%\begin{figure}
%	\centering
%	\begin{subfigure}[b]{0.15\textwidth}
%		\centering
%		\includegraphics[width=1in]{figure/k_1}
%		\caption{$k=1$}
%	\end{subfigure}
%	\begin{subfigure}[b]{0.15\textwidth}
%		\centering
%		\includegraphics[width=1in]{figure/k_2}
%		\caption{$k=2$}
%	\end{subfigure}
%	\begin{subfigure}[b]{0.15\textwidth}
%	\centering
%	\includegraphics[width=1in]{figure/fault_order_n_c}
%	\caption{$k=10$}
%\end{subfigure}
%	\caption{Comparison of robots' configuration and beliefs at the final step with two approaches, $k=1, \,2,\,10$ when two of the sensors fails (circled) and $r_{\textup{eff}}=\infty$, $\Sigma_B = 0.04\mathbf{I}$ for both cases }
%	\label{fig:fig6}
%\end{figure}
%\begin{figure}
%	\centering
%	\includegraphics[width=1.6in]{figure/fault_kl2}
%	\caption{Comparison of K-L divergence of the ground-truth between multiple strategies ($k=1,\,2,\,10$), when two sensors fail during the evolution}
%	\label{fig:fig7}
%\end{figure}

\noindent \textit{Robustness to Sensor Failure:} We next present several examples with varying numbers of sensor failures wherein the robust method clearly illustrates appealing behavior when compared to its less robust counterparts.
In this experiment, the number of sensor failures was varied by $\mathcal{F} \in \{ \lbrace 1 \rbrace,\,\lbrace 1,2\rbrace,\,\lbrace 1,2,3 \rbrace \}$.
Results for robots configuration and target distributions after the $10^{\textup{th}}$ step with the non-robust, robust, and maximum information methods in the case when $\mathcal{F}= \lbrace 1\rbrace$ are shown in Fig. \ref{fig:fig6}. 
Fig. \ref{fig:fig3} shows the time evolutions of root-mean-square error (RMSE) between constructed map and the ground truth map for the non-robust (top) and robust (bottom) methods when $\mathcal{F}=\lbrace 1,2,3 \rbrace$. 
% Notice that RMSE of the robust method is significantly lower than its non-robust counterpart.
Fig. \ref{fig:fig7} compares the K-L divergence between different methods when $\left| \mathcal{F} \right|$ was varied between $1$ and $3$.
As can be seen from Fig \ref{fig:fig6}--\ref{fig:fig7}, the map retrieved by the proposed method when $k=2$ is consistently more robust to sensor failure when compared to existing methods. 
Note in Fig \ref{fig:fig6} (middle and right) the unmapped area is due to the limited sensing range.

\begin{figure*}
	%	\centering
	%	 \begin{subfigure}[b]{0.165\textwidth}
	\centering
	\includegraphics[width=6.9in]{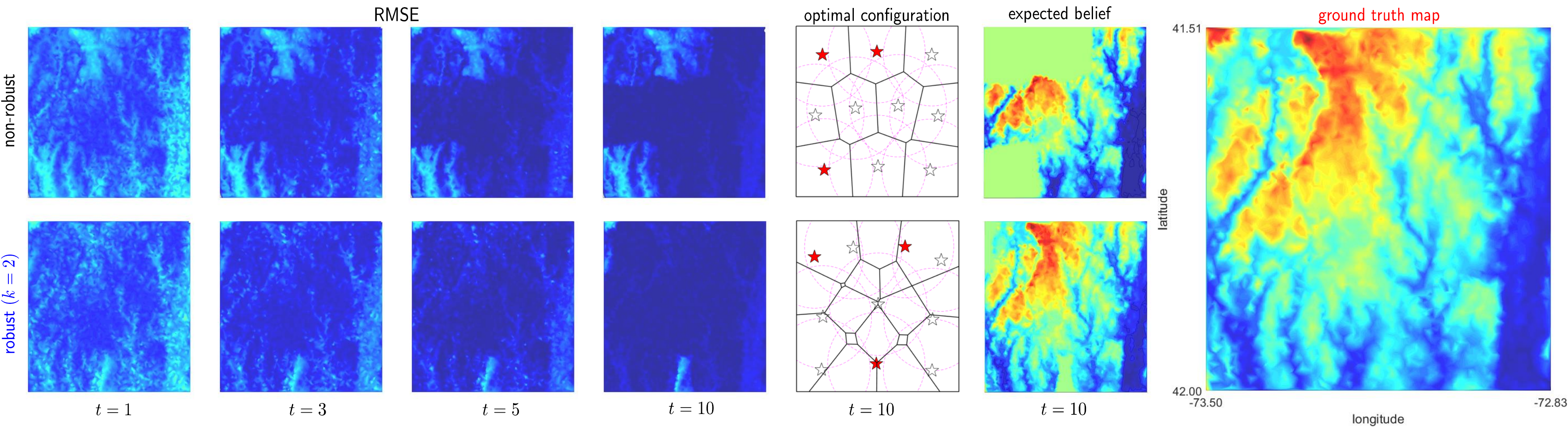}
	%		\caption{$k=1$}
	%	\end{subfigure}
	%	\begin{subfigure}[b]{0.15\textwidth}
	%		\centering
	%		\includegraphics[width=1in]{figure/order2_step_0110_c}
	%		\caption{$k=2$}
	%	\end{subfigure}
	%	 \begin{subfigure}[b]{0.15\textwidth}
	%	\centering
	%	\includegraphics[width=1in]{figure/ordern_step_0110_c}
	%	\caption{$k=10$}
	%\end{subfigure}
	\caption{An illustration of the time evolution of the root-mean-square error (RMSE) (in ft) between the constructed map and the ground truth map for non-robust method (top) and robust method (bottom) where 3 nodes have randomly failed (left 4 images). An illustration of the robot deployment (depicted with stars, where the red stars are failed sensors) sensor footprint (dashed lines), and partition (polygons) for the non-robust (top) and robust method (bottom) is also shown (5th column). The computed expected belief is also depicted from the non-robust (top) and robust (bottom) methods (6th column) and the ground truth image is depicted (right).
	The colormap is using MATLAB's jet colormap.}
% 	when $\mathcal{F}=\lbrace 1,2,3\rbrace$.} 
	%		
	%		 ($k=2$), $r_{\textup{eff}} = 2.89$, $\Sigma_B = 0.04\mathbf{I}$.}
	\label{fig:fig3}
\end{figure*}
\begin{figure}
    \vspace*{-0.1cm}
    \hspace*{-0.3cm}
	\includegraphics[width=1.2in]{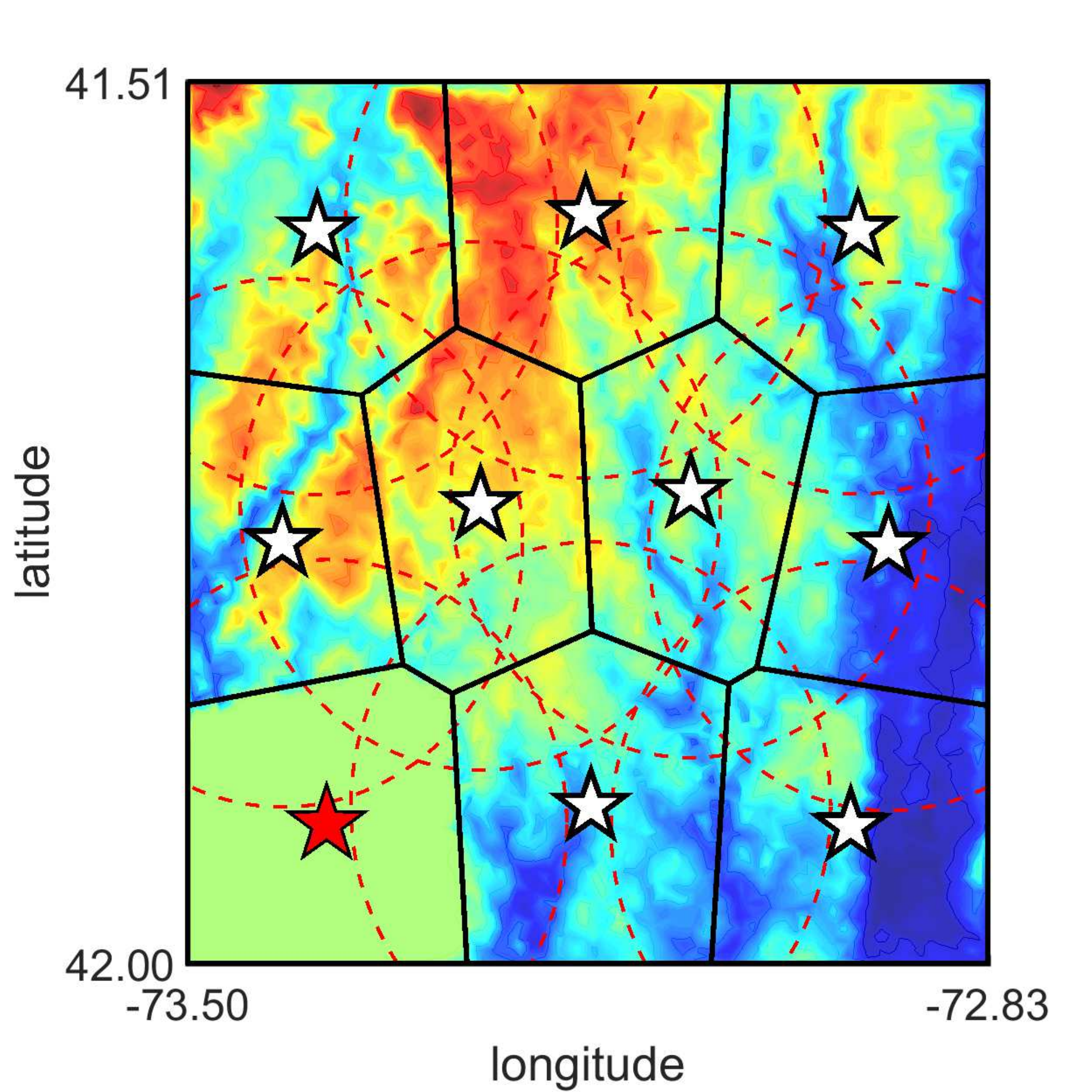}\hspace*{-0.25cm}
	\includegraphics[width=1.2in]{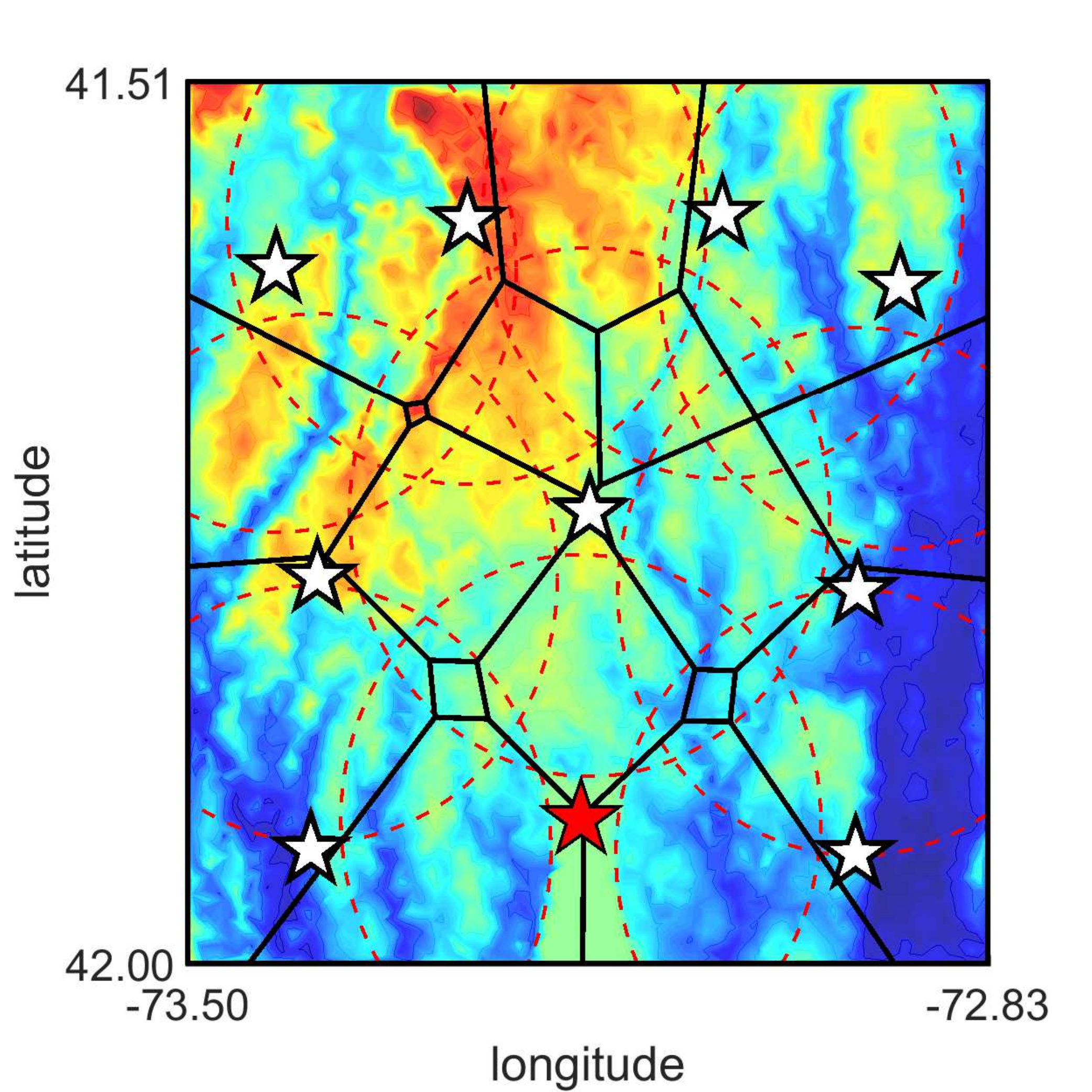}\hspace*{-0.25cm}
	\includegraphics[width=1.2in]{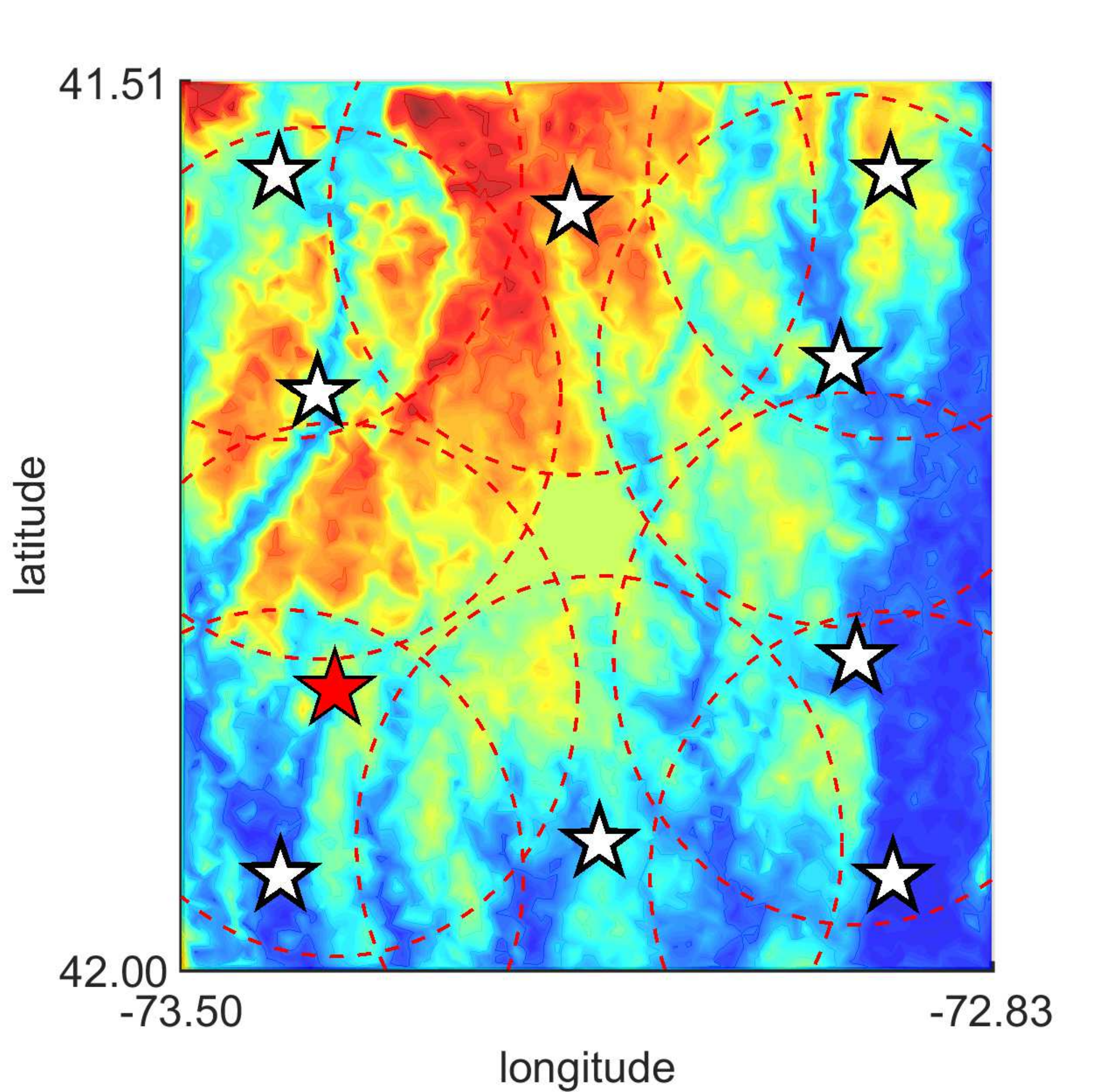}
	\caption{The expected beliefs at $t=10$ from the non-robust method (left), robust method (middle), and max. information gain method (right) along with robot configurations (stars are robot locations and red stars are robots whose sensors have failed), footprints (dashed line), and partition (polygons) is depicted. The ground truth elevation map is on the right of Fig. \ref{fig:fig3}. 
	The colormap is using MATLAB's jet colormap.
	}
	\label{fig:fig6}
\end{figure}
\begin{figure}
\vspace*{-1cm}
	\centering
	\includegraphics[width=3.4in,keepaspectratio=true]{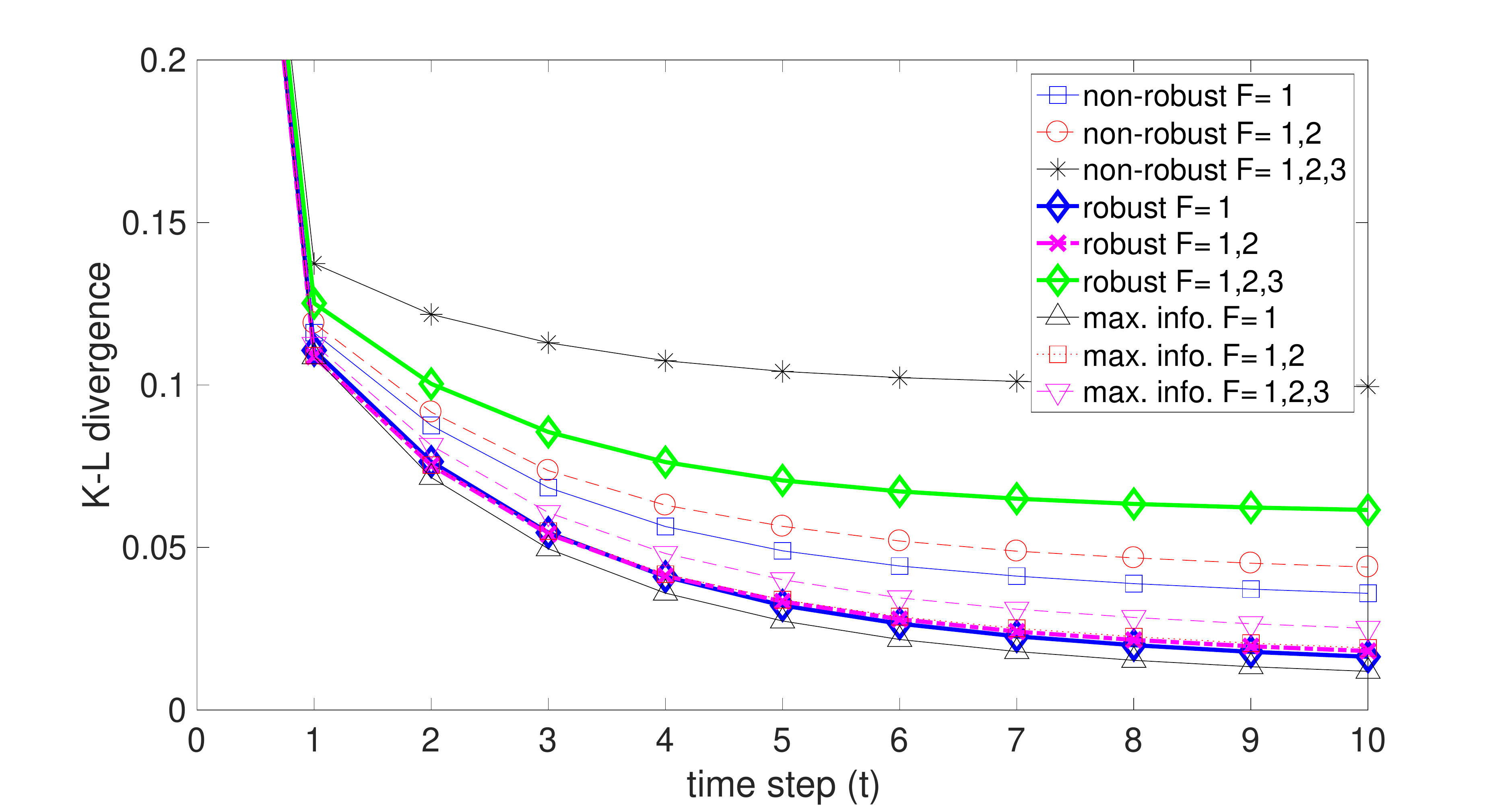}
	\caption{Comparison of K-L divergence of the ground truth distribution between different methods during belief propagation when a component of the nodes have failed.}
	\label{fig:fig7}
\end{figure}
\begin{figure}
\vspace*{-1cm}
	\centering
	\includegraphics[width=3.4in,keepaspectratio=false]{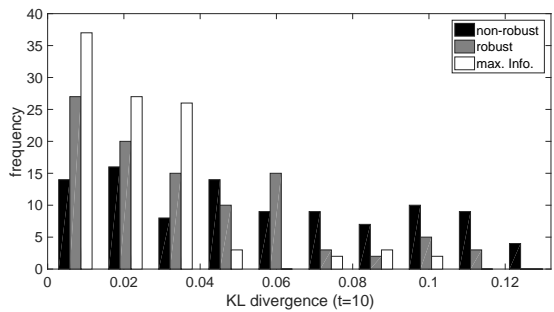}
	\caption{Comparison of robustness between different methods on $100$ randomly generated tests at $t=10$.} 
	\label{fig:fig8}
\end{figure}

\noindent \textit{Statistical Results with Varying Initial Conditions and Fault Compositions:}
Statistical results shows that our method can be used to estimate an arbitrary target distribution given a randomly chosen initial configuration, with different fault compositions well. 
Fig. \ref{fig:fig8} shows a distribution of K-L divergence values at $t=10$ for $100$ test examples consisting of random initial configurations with uniformly sampled number of faults between $1$ and $5$ with $10$ robots.

\noindent \textit{Scalability of our Method:} Fig. \ref{fig:fig0} shows an example with $100$ robots (with $20$ sensors failures) where the robust method outperforms the non-robust method. 
%\noindent \textit{Remark:} It is not surprising to see from the simulation results that the maximum information gain approach has higher robustness against sensor failures than other methods.
This is due to the presence of central information fusion server which requires a full communication throughout the MSN.
This is typically infeasible in real-world applications.

\section{Conclusions and Future Work}
\label{sec:sec9}
This paper presents a deployment strategy for a mobile sensor network to enable the recovery of an environmental map over a bounded space in a manner that is robust to sensor failures. 
% It is expected that our method will become ineffective if there is not enough number of mobile agents having sufficiently long effective sensing ranges relative to the workspace size. 
We plan to employ multi-agent patrolling \cite{portugal2011survey,kemna2017multi} or sweep coverage \cite{rekleitis2004limited} to resolve problems associated with not having enough sensors to fully cover a target space.
Also, as reported in the literature \cite{anguelov2004detecting}, our combined sensor model has been adopted to emulate the real-world laser scanner's behavior; nevertheless, we plan to conduct extensive real world multi-robot experiments for further validation of our range sensor model.
% Lastly, we assumed in this study that the beliefs are shared between robots such that both tasks of information gathering, propagating and approximating belief require a central entity. 
% In the future, we will explore how to devise distributed communication protocol to enable distributed belief estimation. 
%\section{type of sensors}

\bibliographystyle{IEEEtran}
\bibliography{IEEEabrv,reference_park_17}

\end{document}